\documentclass[aps,floats,nofootinbib]{revtex4}

\usepackage{graphicx}
\usepackage{amsmath}

\def\lsim{\mathrel{\raise.3ex\hbox{$<$\kern-.75em\lower1ex\hbox{$\sim$}}}}
\def\gsim{\mathrel{\raise.3ex\hbox{$>$\kern-.75em\lower1ex\hbox{$\sim$}}}}

\def\ie{{\it i.e.\;}}
\def\eg{{\it e.g.\;}}

\providecommand{\SP}{\scriptscriptstyle}

\newcommand{\mst}[1]{m_{\tilde{t}_{\SP {#1}}}}
\newcommand{\mhp}{m_{H^\pm}}

\begin{document}
\hfill$\vcenter{\hbox{\bf hep-ph/0506061}}$

\vskip 0.5cm

\title {Dark Matter and Collider Phenomenology with two light
Supersymmetric Higgs Bosons}

\author{Dan Hooper$^1$ and Tilman Plehn$^2$} 
\affiliation{$^1$ University of Oxford, Oxford, United Kingdom;\\
             $^2$ Max Planck Institute for Physics, Munich, Germany}

\date{\today} 

\bigskip

\begin{abstract}

Recently, it has been pointed out that two different excesses of events
observed at LEP could be interpreted as the CP-even Higgs bosons of the MSSM
with masses of approximately 98 and 114~GeV. If this is the case, the entire
MSSM Higgs sector is required to be light. In this article, we explore such a
scenario in detail. We constrain the Higgs and supersymmetric spectrum using
$B$ physics constraints as well as the magnetic moment of the muon. We then
point out the implications for neutralino dark matter --- next generation
direct detection experiments will be sensitive to all MSSM models with such a
Higgs sector. Finally, we find that all models outside of a very narrow
corridor of the parameter space have a charged Higgs boson which will be
observed at the LHC. In those exceptional models which do not contain an
observable charged Higgs, a light top squark will always be seen at the LHC,
and likely at the Tevatron.

\end{abstract}

\pacs{PAC numbers: 14.80.Cp, 14.80.Ly, 12.60.Jv, 95.35.+d}
\maketitle 

\section{Setting the Stage}\label{sec:intro}

The four experiments at LEP have searched
for Higgs bosons up to a mass of approximately 115
GeV~\cite{LEP}. Although no strong indication of a Higgs has been
detected by LEP, excesses with statistical significances of
2.3$\sigma$ and 1.7$\sigma$ have been reported for Higgs-like events
corresponding to masses of 98~GeV and 115~GeV, respectively. Very
recently, it has been pointed out by Manuel Drees that both of these
excesses reported by LEP could be accommodated within the Minimal
Supersymmetric Standard Model (MSSM) as signatures of the two CP-even
Higgs bosons~\cite{drees}.

One of the reason that the LEP excess at 98~GeV has received so little
attention might be that the rate observed corresponding to this mass is about
a factor of ten below that which would be expected from a Standard Model Higgs
boson. In the MSSM, however, the $h$-$Z$-$Z$ coupling is suppressed relative
to the value in the Standard Model by a factor of $\sin(\beta-\alpha)$, where
$\alpha$ is the mixing angle between the two CP-even Higgs bosons and $\tan
\beta$ is the ratio of the two vacuum expectation values. To predict the
number of events seen at LEP for a 98~GeV light Higgs boson, we constrain
$0.056 \lsim \sin^2(\beta-\alpha) \lsim 0.144$ (the precise range of
$\sin^2(\beta-\alpha)$, $m_h$ and $m_H$ we allow is the same as used in
Ref.~\cite{drees} to allow for easy comparison of results).  Meanwhile, the analogous
coupling of the heavy CP-even Higgs is only slightly suppressed below the
Standard Model value and is consistent with the number of events reported by
LEP corresponding to a 115~GeV Higgs boson. These two excesses together yield
a combined statistical significance of 3.1$\sigma$~\cite{drees}.\medskip

The mass scale of the MSSM Higgs sector is usually fixed by the mass
of the pseudoscalar Higgs boson $m_A$. In the well--known leading
$m_t^4$ approximation~\cite{mt_approx,carlos}, we can very roughly link the
pseudoscalar mass to the two scalar Higgs masses in the limit of
non-mixing top squarks:
\begin{alignat}{4}
m_h^2 + m_H^2 &=&& \; m_A^2 + m_Z^2 + \epsilon \notag \\
\mhp^2        &=&& \; m_A^2 + m_W^2 \notag \\
\epsilon      &=&& \; \frac{3 G_F}{\sqrt{2} \pi^2}
                   \; \frac{m_t^4}{\sin^2\beta}
                   \; \log \frac{\mst{}^2}{m_t^2}
\label{eq:mt_approx}
\end{alignat}
In this approximation $\mst{}$ is the top squark mass, $m_{h,A,H}$ are the
neutral Higgs masses and $\mhp$ is the charged Higgs mass.  We
immediately see that the existence of two light scalar Higgs bosons leads to all of the MSSM Higgs
bosons being fairly light. This can lead to interesting phenomenology in the MSSM. In this article, we consider the constraints on
this scenario from the $B \to X_s \gamma$ and $B_s \to \mu^+ \mu^-$ branching
fractions and the magnetic moment of the muon.  We discuss the implications of
such a model on the phenomenology of neutralino dark matter, and find that the
prospects for the direct detection of neutralinos in elastic scattering
experiments to be excellent. We then study the spectra of the superpartners
and the Higgs sector with two light Higgs scalars and discuss in detail the
prospects for discovering the charged Higgs boson at the LHC. \bigskip

\begin{figure}[t]
\includegraphics[width=2.0in,angle=90]{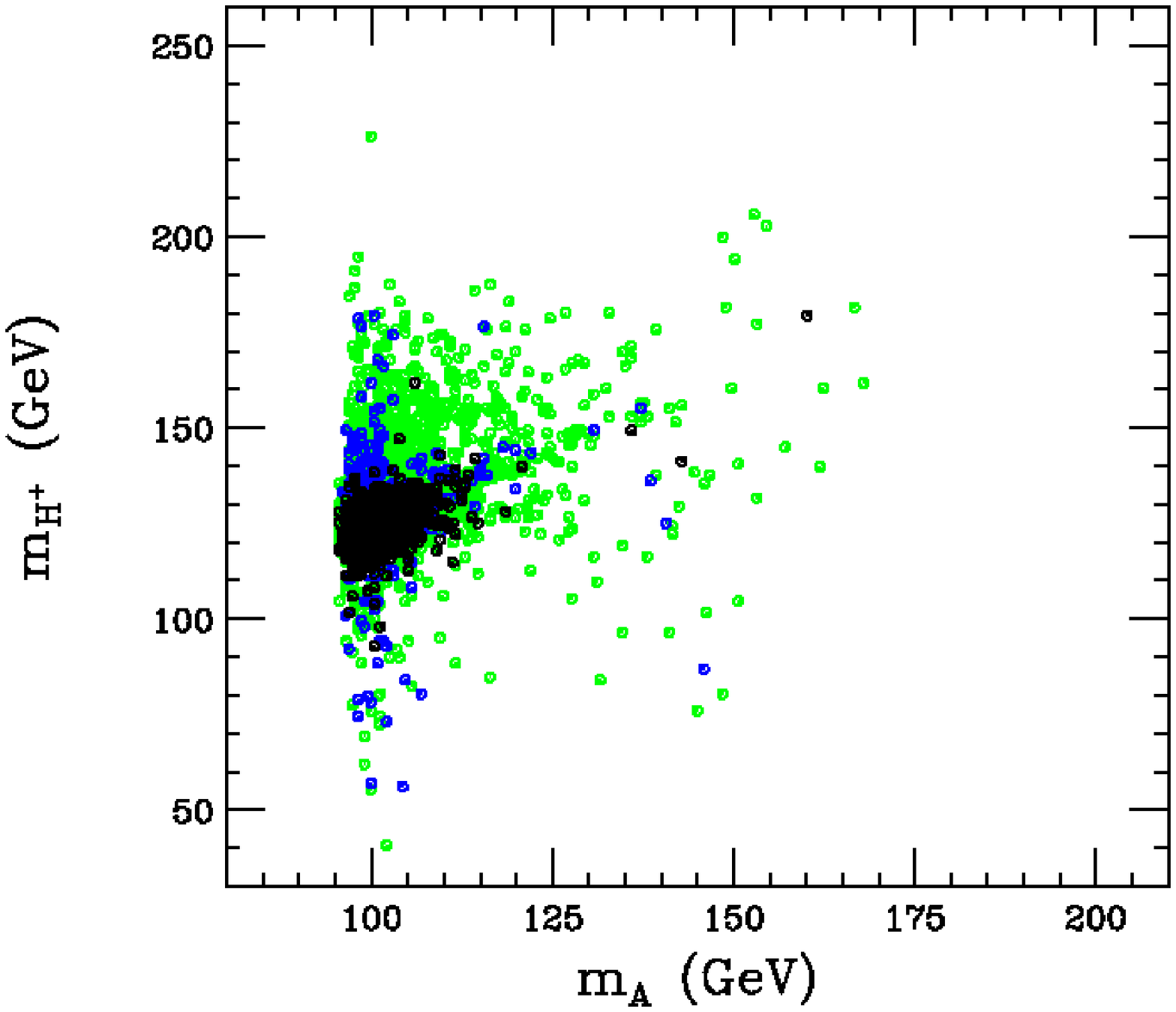} \hspace*{30mm}
\includegraphics[width=2.0in,angle=90]{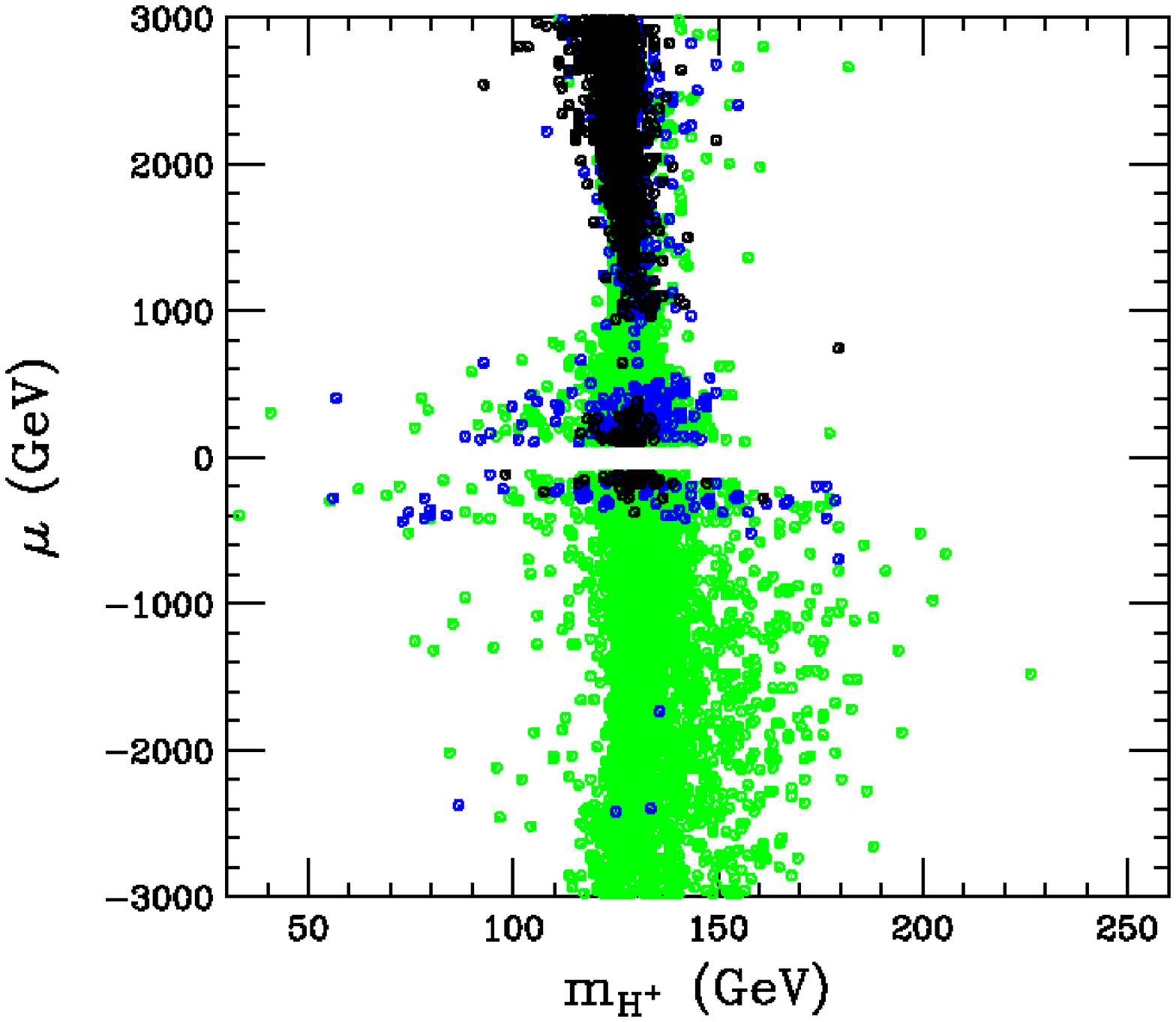}
\caption{The mass of the pseudoscalar Higgs versus the charged Higgs is
shown in the left frame. In the right frame, $\mu$ is shown with 
the charged Higgs mass. For all points shown, $95 \, \rm{GeV} < m_h <
101 \, \rm{GeV}$, $111 \, \rm{GeV} < m_H < 119 \, \rm{GeV}$ and $0.056
\lsim \sin^2(\beta-\alpha) \lsim 0.144$, corresponding to the range
matching the observations at LEP. The black points are consistent with
measurements of $B \to X_s \gamma$ at the 3$\sigma$ level and
do not violate the Tevatron constraint on the $B_s \to \mu^+ \mu^-$
branching fraction. Blue points violate $B_s \to \mu^+
\mu^-$, but are consistent with $B \to X_s \gamma$. Green
points violate $B \to X_s \gamma$. These constraints will be
discussed further in section~\ref{sec:constraints}.}
\label{fig:achmuc}
\end{figure}

In order to study the phenomenology of supersymmetric models with light
Higgs scalars, we first performed a scan over the relevant parameters of the
MSSM. We have varied all masses up to 6~TeV, and $\tan \beta$ between 1 and 60. We did
not assume any specific supersymmetry breaking scenario or unification scheme.
To take into account the radiative corrections to the Higgs masses, we have used
FeynHiggs~\cite{feynhiggs}, which includes all contributions up to the
two-loop level. We find that this level of precision is needed to obtain an
accurate representation of the MSSM phenomenology within this class of models.
We give a detailed analysis of the Higgs sector in
Section~\ref{sec:coll}.\smallskip

In Fig.~\ref{fig:achmuc}, we confirm the conclusion of Ref.~\cite{drees} that
the entire MSSM Higgs sector is required to be rather light to accommodate
both excesses reported by LEP. In the overwhelming majority of models found,
the mass of the pseudoscalar Higgs bosons is between 95 and 130~GeV and the
mass of the charged Higgs lies between 110 and 150~GeV. Although we do find
some points outside of this range, in particular a trail of models extending
to the upper right of the figure (left frame), these are quite rare compared to the dense
collection of models near $m_A \sim 110$~GeV and $m_{H^{\pm}} \sim 130$~GeV.
For each point shown, we require $95 \, \rm{GeV} < m_h < 101 \, \rm{GeV}$,
$111 \, \rm{GeV} < m_H < 119 \, \rm{GeV}$ and $0.056 \lsim
\sin^2(\beta-\alpha) \lsim 0.144$, corresponding to the range needed to match
the observations at LEP. The differences between this figure and the
corresponding one in Ref.~\cite{drees} come from the fact that we scan continuously over
$\tan\beta$. In the right frame of Fig.~\ref{fig:achmuc}, we check the mass of
the charged Higgs boson for correlations with the value of $\mu$. We have
checked that indeed the trails of scattered points at large pseudoscalar
masses correspond to the larger values of $\tan\beta$ the larger the
corresponding charged Higgs masses become. Without imposing
additional constraints we find relatively little correlation, for example,
between the Higgs masses and the $\mu$ parameter, apart from the chargino mass
limit of 104~GeV from LEP.

\section{\boldmath Indirect Constraints from $B \to X_s \gamma$, $B_s \to \mu^+ \mu^-$ and $(g-2)_{\mu}$}
\label{sec:constraints}

\begin{figure}[t]
\includegraphics[width=2.0in,angle=90]{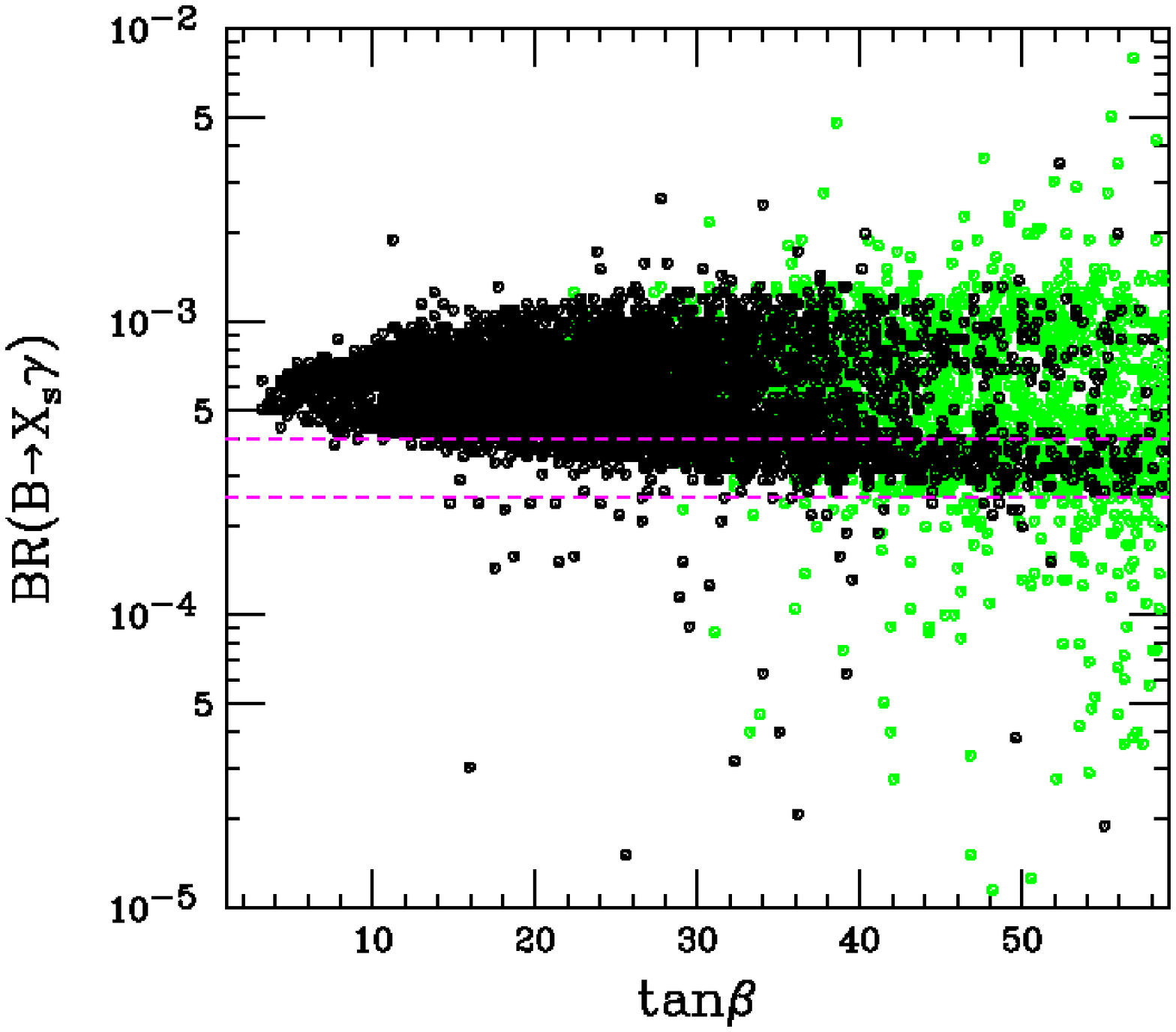}
\includegraphics[width=2.0in,angle=90]{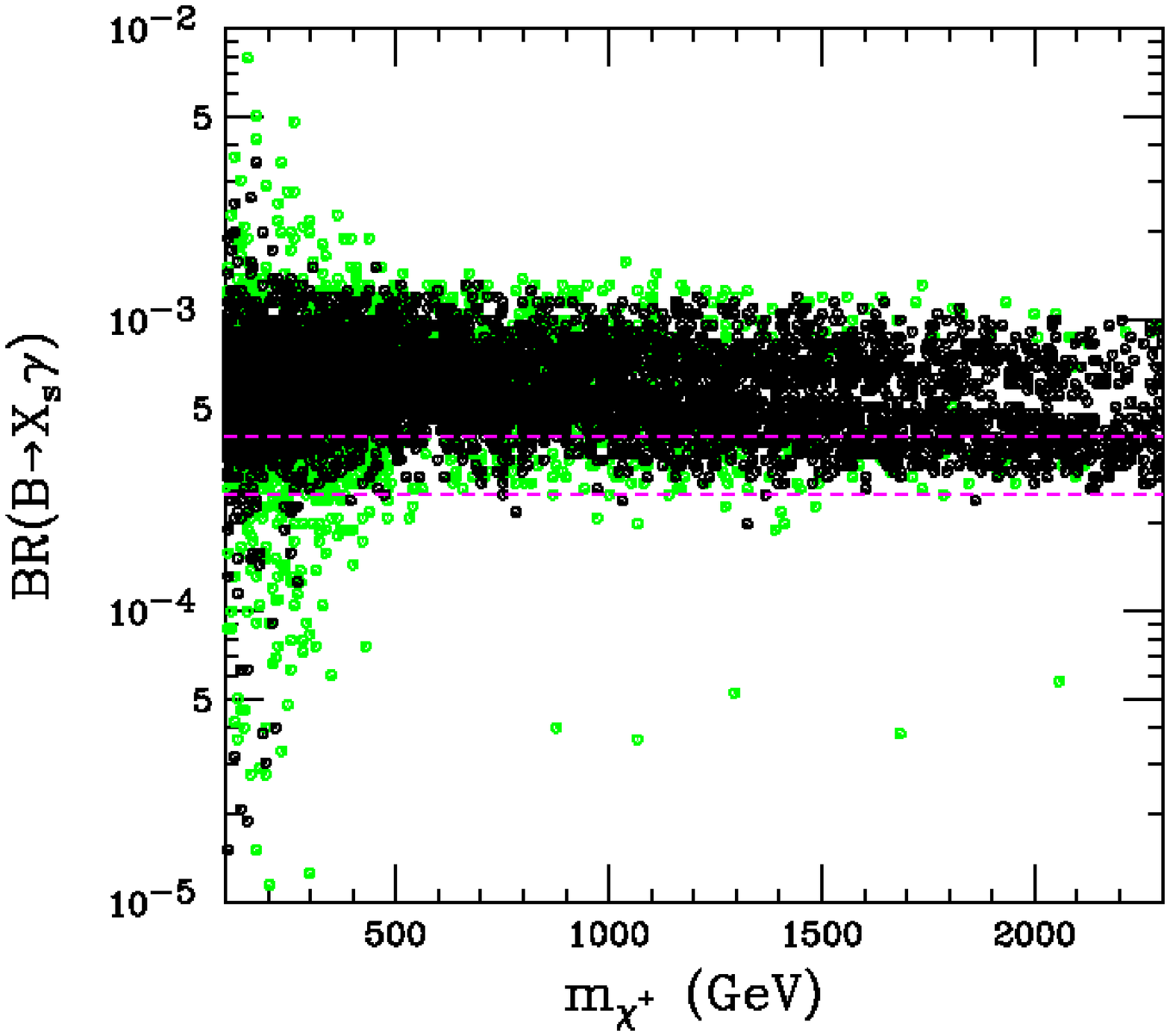}
\includegraphics[width=2.0in,angle=90]{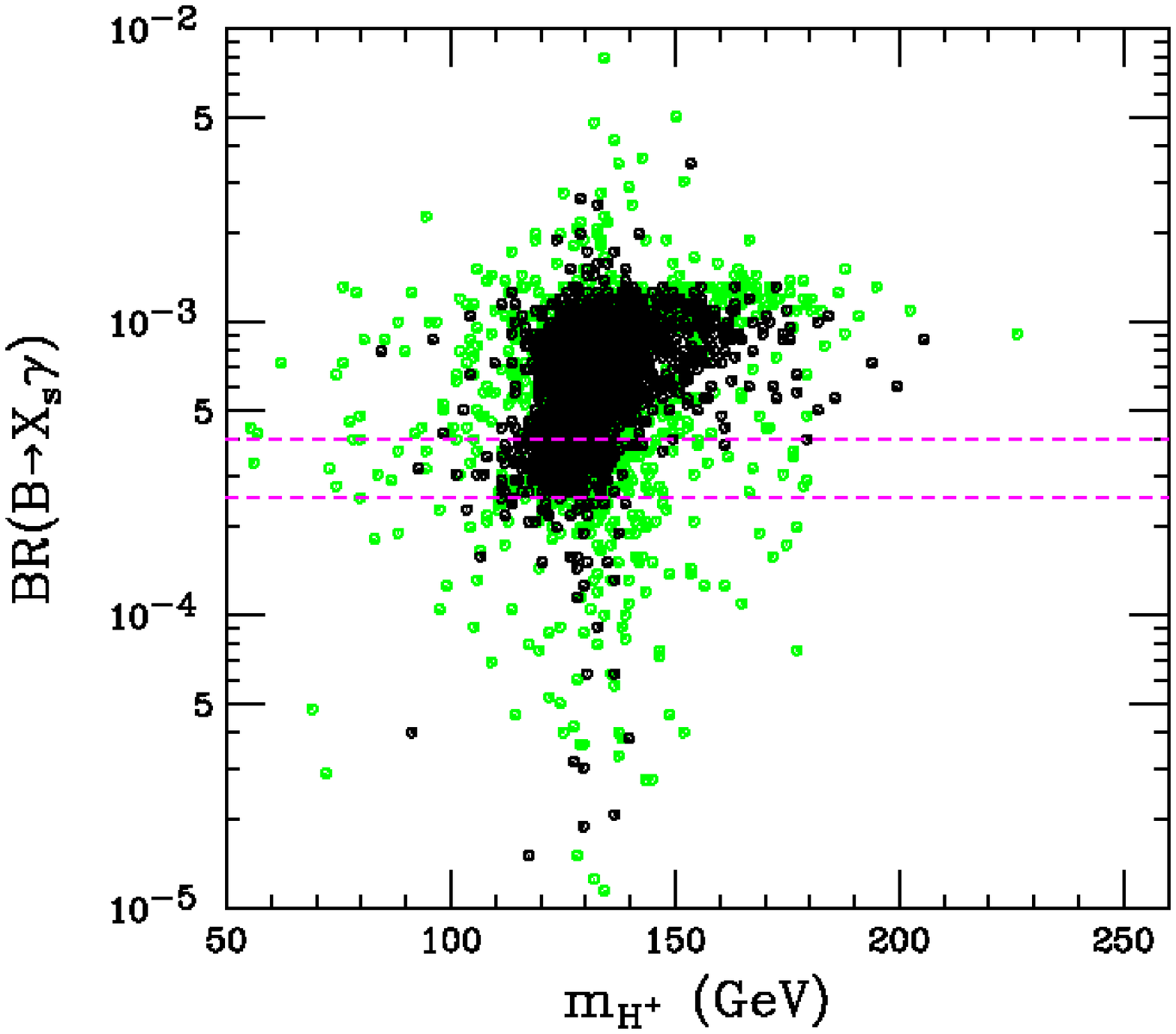}
\caption{The $b \to s \gamma$ branching fraction as a function
of $\tan \beta$, the lightest chargino mass, and the charged Higgs
mass. Shown as horizontal dashed lines are the $2 \sigma$ confidence
bounds on this quantity as measured by BELLE, CLEO and ALEPH. Again,
all points shown have $95 \, \rm{GeV} < m_h < 101 \, \rm{GeV}$, $111
\, \rm{GeV} < m_H < 119 \, \rm{GeV}$ and $0.056 \lsim
\sin^2(\beta-\alpha) \lsim 0.144$. Black points do not violate the Tevatron
constraint on the $B_s \to \mu^+ \mu^-$ branching
fraction. Green points violate this constraint.}
\label{fig:bsg}
\end{figure}

Constraints from rare $B$-decays and the magnetic moment of the muon
are some of the most useful tools we currently have to guide our studies of
supersymmetric phenomenology. The $B$ physics constraints gain their
power mostly because of a possible $\tan\beta$ enhancement of the
bottom Yukawa couplings. In this section, we consider some of these
constraints within the context of models with 98 and 115~GeV Higgs
bosons.\smallskip

The branching fraction of $B$ decays to a strange state plus a photon
has been measured by the BELLE~\cite{belle}, CLEO~\cite{cleo} and
ALEPH~\cite{aleph} experiments. The weighted average of these
experiments' results indicate BR$(B \to X_s \gamma)= (3.25 \pm
0.37) \times 10^{-4}$. In comparison, the Standard Model prediction
for this transition rate is $(3.70 \pm 0.30) \times
10^{-4}$~\cite{bsa_theo}.

The supersymmetric processes which are most likely to contribute substantially
to this branching ratio involve a charged Higgs or a chargino. These
contributions are enhanced by powers of the Yukawa coupling $m_b \tan\beta$
for large values of $\tan \beta$. In passing we emphasize that there are
additional $\Delta_b$ corrections~\cite{niere} which can have huge effects if
the MSSM spectrum is split between light gluinos or higgsinos and heavier
sbottoms, but we will see that this is not the part of parameter space in which
we are interested in. If the gluino mass is smaller than 300~GeV, the LHC will
be swamped by gluino pair production with cross sections as large as
$1000$~pb and the effect of the gluino mass on the charged Higgs boson is
negligible.\smallskip

In Fig.~\ref{fig:bsg}, we plot BR$(B \to X_s \gamma)$ versus $\tan \beta$
(left frame), the lightest chargino mass (center frame) and the charged Higgs
mass (right frame). Shown as horizontal dashed lines are the $2 \sigma$
confidence bounds on this branching fraction. Again, all points shown have $95
\, \rm{GeV} < m_h < 101 \, \rm{GeV}$, $111 \, \rm{GeV} < m_H < 119 \,
\rm{GeV}$ and $0.056 \lsim \sin^2(\beta-\alpha) \lsim 0.144$. Strictly
applying a $2\sigma$ limit would rule out all points with $\tan\beta \lesssim
10$, but on the other hand allowing for a $3 \sigma$ deviation from the
Standard Model brings all values of $\tan\beta$ back into the allowed region.
Note that the preference of larger values of $\tan\beta$ is due to a slight
bias toward a finite MSSM contribution: if we force the Higgs sector to be
light, the 2HDM diagrams will lead to an increase of the observable BR$(B \to
X_s \gamma)$ by typically tens of percent up to factor of two. This charged
Higgs contributions consist of a $\tan\beta$ suppressed term and a constant
term, but does not exhibit any $\tan\beta$ enhancement (if we do not consider
anomalously large gluino loops). Because the measured value of BR$(B \to X_s
\gamma)$ is actually slightly smaller than the Standard Model prediction, the
chargino has to compensate for the 2HDM contribution. Because the chargino
contribution to BR$(B \to X_s \gamma)$ is enhanced by one power of $\tan\beta$
on the amplitude level, we can achieve this by choosing large values of
$\tan\beta$, as we see in the left panel of Fig.~\ref{fig:bsg}. We can, of
course, try to increase the chargino diagram by making the chargino light, but
this is much less efficient, because the loop involved is a chargino--stop
loop, so that just making the chargino light has comparably little effect.
Again, we see in the center frame of Fig.~\ref{fig:bsg} that very light
chargino masses serve this purpose. Moreover, the large chargino contribution
has to come with the right sign and therefore prefers positive $\mu$, as is shown in Fig.~\ref{fig:achmuc}. As we can see, the charged Higgs mass and
therefore the contribution to BR$(B \to X_s\gamma)$ is
basically fixed by the two light Higgs masses. Thus merely shifting
the charged Higgs mass between 120 and 170~GeV has even less of an effect.\medskip

\begin{figure}[t]
\includegraphics[width=2.0in,angle=90]{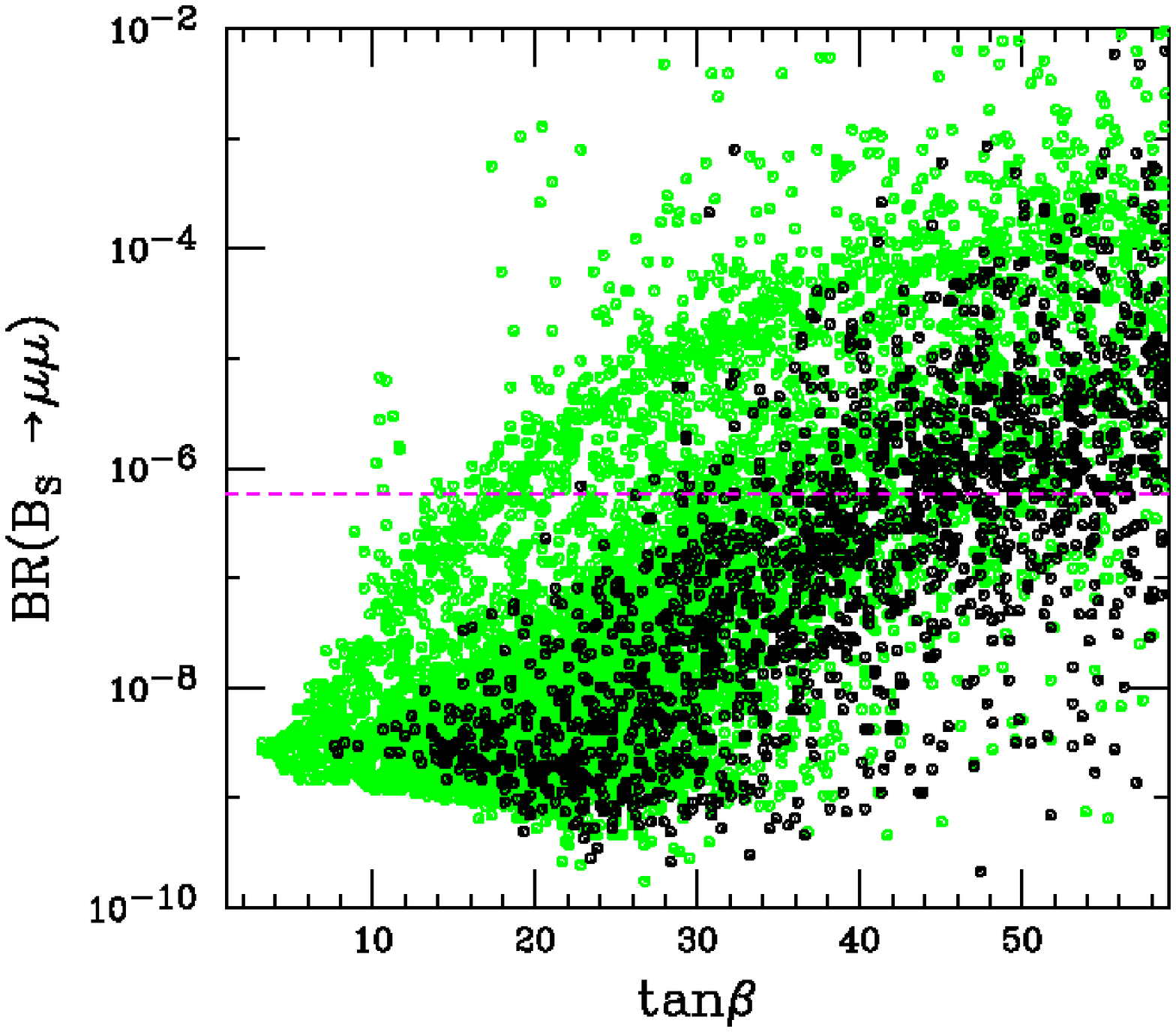}
\includegraphics[width=2.0in,angle=90]{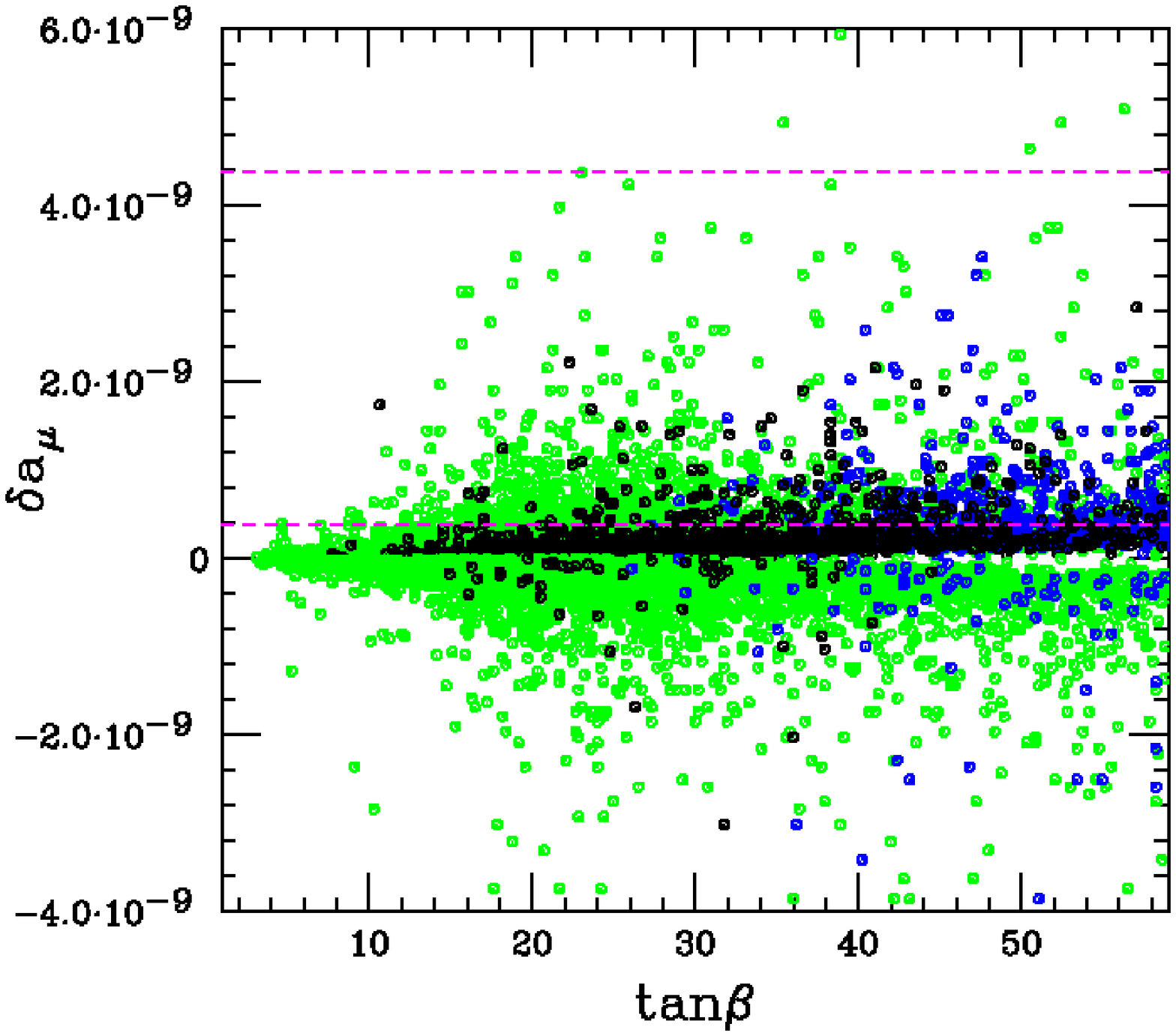}
\includegraphics[width=2.0in,angle=90]{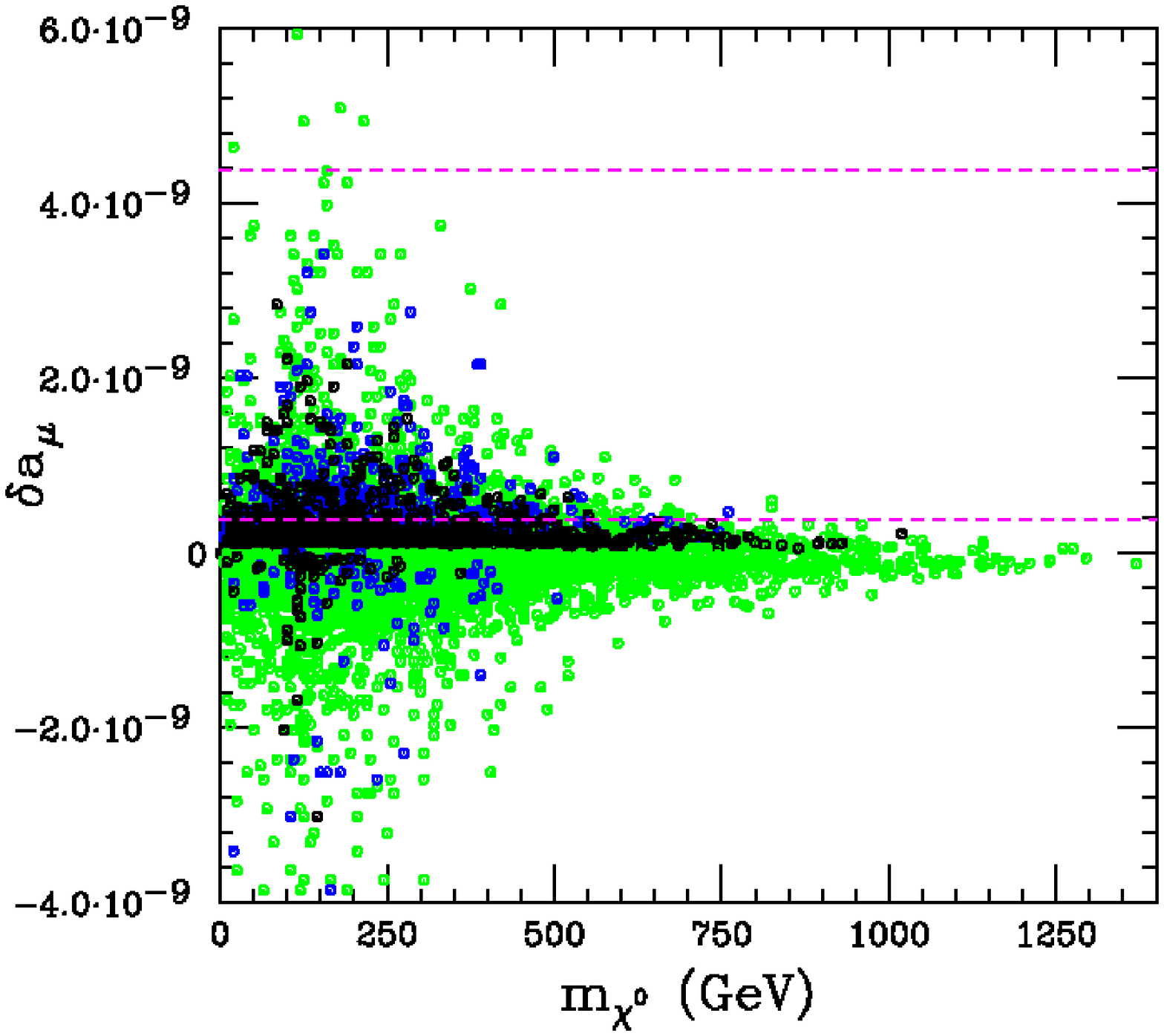} 
\caption{Left: The $B_s \to \mu^+ \mu^-$ branching fraction versus
$\tan \beta$. The dashed horizontal line is the 90\% upper confidence
bound placed on this branching fraction by the Tevatron experiments. Black points
are consistent with measurements of $B \to X_s \gamma$ at the
3$\sigma$ level. Green points violate this constraint. Right two
panels: the contribution to the magnetic moment of the muon. The
dashed horizontal lines are the $2\sigma$ bounds from $e^+ e^-$
data. The black points are consistent with measurements of $B \to X_s
\gamma$ at the 3$\sigma$ level and do not violate the Tevatron constraint
on the $B_s \to \mu^+ \mu^-$ branching fraction. Blue points do
violate $B_s \to \mu^+ \mu^-$, but are consistent with $B \to X_s
\gamma$. Green points violate $B \to X_s \gamma$. All points shown
have $95 \, \rm{GeV} < m_h < 101 \, \rm{GeV}$, $111 \, \rm{GeV} < m_H
< 119 \, \rm{GeV}$ and $0.056 \lsim \sin^2(\beta-\alpha) \lsim
0.144$.}
\label{fig:bmumu}
\label{fig:g2}
\end{figure}

Limits on the branching fraction of the rare decay $B_s \to \mu^+
\mu^-$ can also be exploited to limit the allowed MSSM parameter
space. The Standard Model prediction for this branching ratio is
BR=$(2.4 \pm 0.5) \times 10^{-9}$. In particular, the Higgs--induced
MSSM contributions to this branching fraction scales with $\tan^6
\beta/m_A^4$ and tends to be especially large in our scenario with a
light Higgs sector~\cite{bmumu_theo}. If we require the Higgs sector to be light, models
with very large values of $\tan \beta$ are likely to violate the
constraints on this quantity placed at the Tevatron~\cite{bmumu_tev}. In
Fig.~\ref{fig:bmumu}, we plot BR$(B_s \to \mu^+ \mu^-)$ versus $\tan
\beta$. Shown as a dashed horizontal line is the 90\% upper confidence
bound placed by the Tevatron experiments. The Standard Model value is clearly visible as
the low $\tan\beta$ nose in the distribution of the MSSM parameter
points --- small $\tan\beta$ means small contributions from the MSSM
Higgs sector. The upper limit from the Tevatron becomes an issue for
$\tan\beta>10$ but it really only develops seriously destructive power for
$\tan\beta>50$. \medskip

Combining the upper limit on the $B_s \rightarrow \mu^+ \mu^-$ decay and the measurement of $B \to
s\gamma$ makes it considerably harder for us to find viable MSSM
scenarios with two light Higgs scalars. On the one hand, we
need an MSSM contribution to $B \to X_s \gamma$ from the chargino sector with
the right sign and magnitude to compensate for the charged Higgs diagrams.
This is possible by exploiting the enhancement by one power of $\tan\beta$ in
the chargino amplitude. On the other hand, we do not want too large a
contribution to $B_s \to \mu^+ \mu^-$ from the same sector which scales as $\tan^6\beta$. Putting both of these constraints together, we favor fairly large, but not too large, values of
$\tan\beta$ and at the same time accept some light superpartners to accommodate $B
\to X_s \gamma$. We should stress, however, that this argument is not strict in
the sense that these two constraints guarantee a light MSSM mass spectrum.
By choosing properly tuned parameters, we can still get by with a spectrum where the lightest of the charginos and stops weighs more than 1~TeV.

Looking into the near future, we see that of course the LHC prospects of seeing
these light-Higgs MSSM scenarios in $B$ physics are excellent. Simulations of
CMS, ATLAS and LHCb events probing BR$(B_s \to \mu^+ \mu^-)$ predict
tens of events for 10~fb~${}^{-1}$ and the Standard Model decay rate.
Moreover, it has been shown that high luminosity triggering on this search
channel is possible, so that the LHC reach should even cover smaller branching
fractions than predicted in the Standard Model. Basically all parameter points
shown in Fig.~\ref{fig:bmumu} will be clearly visible~\cite{bmumu_lhc}. The
only question is whether the theoretical and experimental errors will allow us to distinguish between the Standard Model and the MSSM predictions.\medskip

Finally, the magnetic moment of the muon has been measured to be anomalously
high in comparison to the Standard Model prediction. Using $e^+ e^-$ data, the
measured value exceeds the theoretical prediction by $\delta a_\mu
(e^+e^-)=23.9 \pm 7.2_{\rm{had-lo}} \pm 3.5_{\rm{lbl}} \pm 6_{\rm{exp}} \times
10^{-10}$, where the error bars correspond to theoretical uncertainties in the
leading order hadronic and the hadronic light-by-light contributions as well
as from experimental contributions~\cite{gminus2data,gminus2sm}. This measured value is 2.4$\sigma$ above
the Standard Model prediction. Experiments using $\tau$ data, on the other
hand, find $\delta a_\mu (\tau^+ \tau^-)=7.6 \pm 5.8_{\rm{had-lo}} \pm
3.5_{\rm{lbl}} \pm 6_{\rm{exp}} \times 10^{-10}$, which is only 0.9$\sigma$
above the Standard Model prediction and likely not in agreement with most
recent KLOE data~\cite{gminus2kloe}. Given this conflict and the marginal statistical significance of these measurements, we do not require all of our scenarios to produce the measured value of $(g-2)_\mu$.  

Contributions to $(g-2)_{\mu}$ occur at the one and two loop levels from
diagrams involving both Higgs bosons~\cite{gminus2higgs} and
superpartners~\cite{gminus2susy}. At the one-loop level the contribution
from superpartner exchange is proportional to $a_\mu^{\rm SUSY} \propto
\tan\beta/m_{\rm SUSY} \; {\rm sign}(\mu)$~\cite{gminus2susy}. The exchange of a
light pseudoscalar Higgs contributes like $a_\mu^{\rm 2HDM} \propto
\tan^2\beta/m_A^2$ in the leading order of $\tan\beta$. Because we are only
considering models with a light Higgs sector, the effects from the $A$
exchange will dominate. This is what we see in the $\tan\beta$ dependence of
the permitted parameters points in Fig.~\ref{fig:g2}. In a way, the situation
is the same as for $B \to X_s \gamma$: both measurements sit very slightly away from their respective Standard Model predictions (and from a statistical point of view are not very convincing), and the central values can be accommodated by
choosing large $\tan\beta$. In the two right panels of Fig.~\ref{fig:g2} we
see that the $B \to X_s\gamma$ constraint goes a long way to also accommodate the $(g-2)_\mu$ measurement in our MSSM parameter space. This is particularly striking when we
look at the behavior of the points around $\delta a_\mu = 0$. The $B \to
X_s\gamma$ constraint disfavors the `wrong' sign of $\delta a_\mu$ already, so
that the impact of the $(g-2)_\mu$ measurement on our light-Higgs models is
very limited once we allow slightly more than a $2\sigma$ window.  Again, we
see how very large values of $\tan\beta$ are disfavored by $B_s \to \mu^+
\mu^-$, which drives the MSSM parameter points toward lighter superpartners,
\ie lighter stops and charginos/neutralinos. This constraint is expected to
become more stringent over the coming years, even before the LHC will start
operation.\smallskip


To calculate BR$(B \to X_s \gamma)$, BR$(B_s \to \mu^+
\mu^-)$ and $\delta a_{\mu}$, we have used the micrOMEGAs
program~\cite{micromegas}.


\section{Implications for Neutralino Dark Matter}

\begin{figure}[t]
\includegraphics[width=2.0in,angle=90]{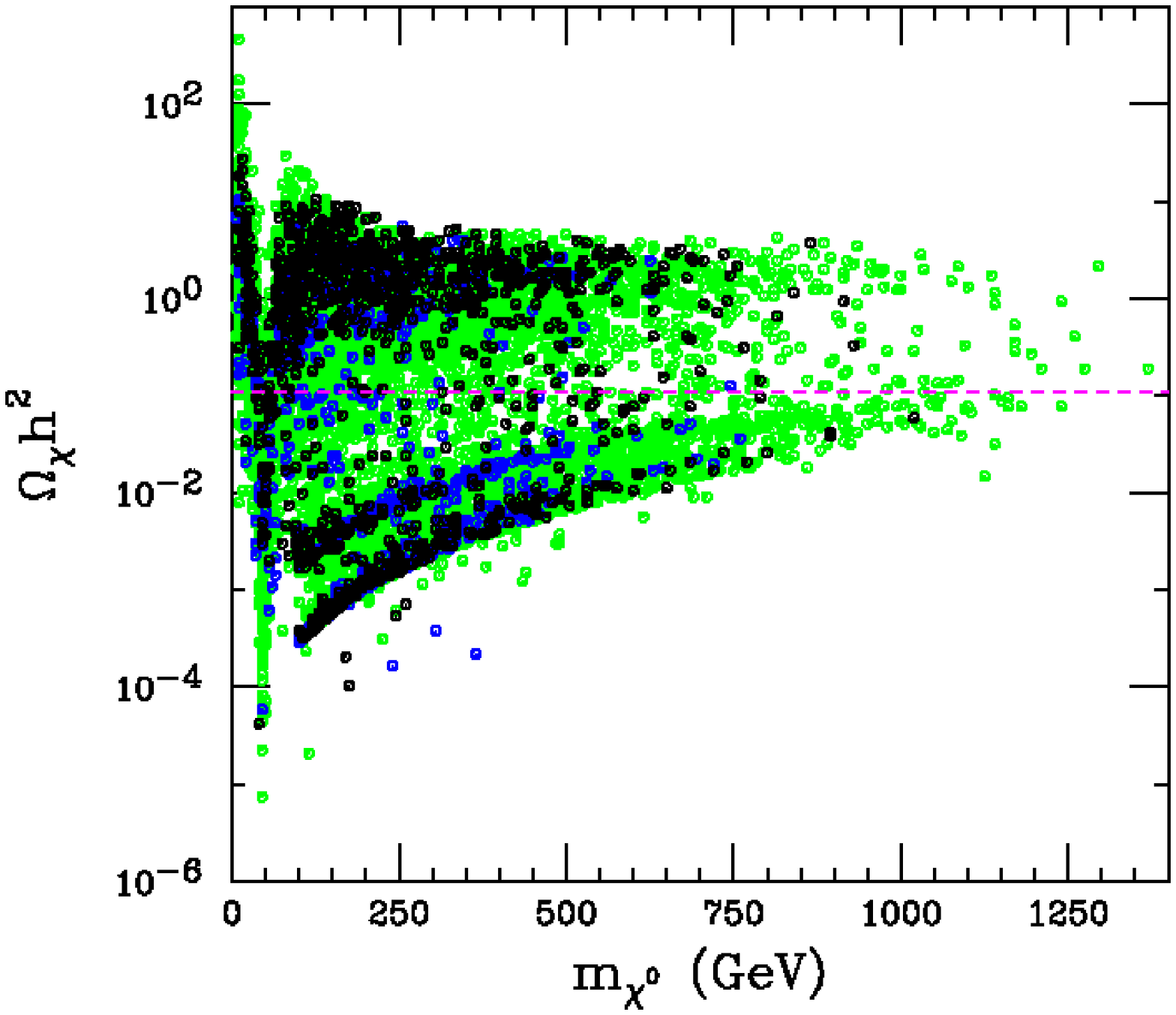} \hspace*{30mm}
\includegraphics[width=2.0in,angle=90]{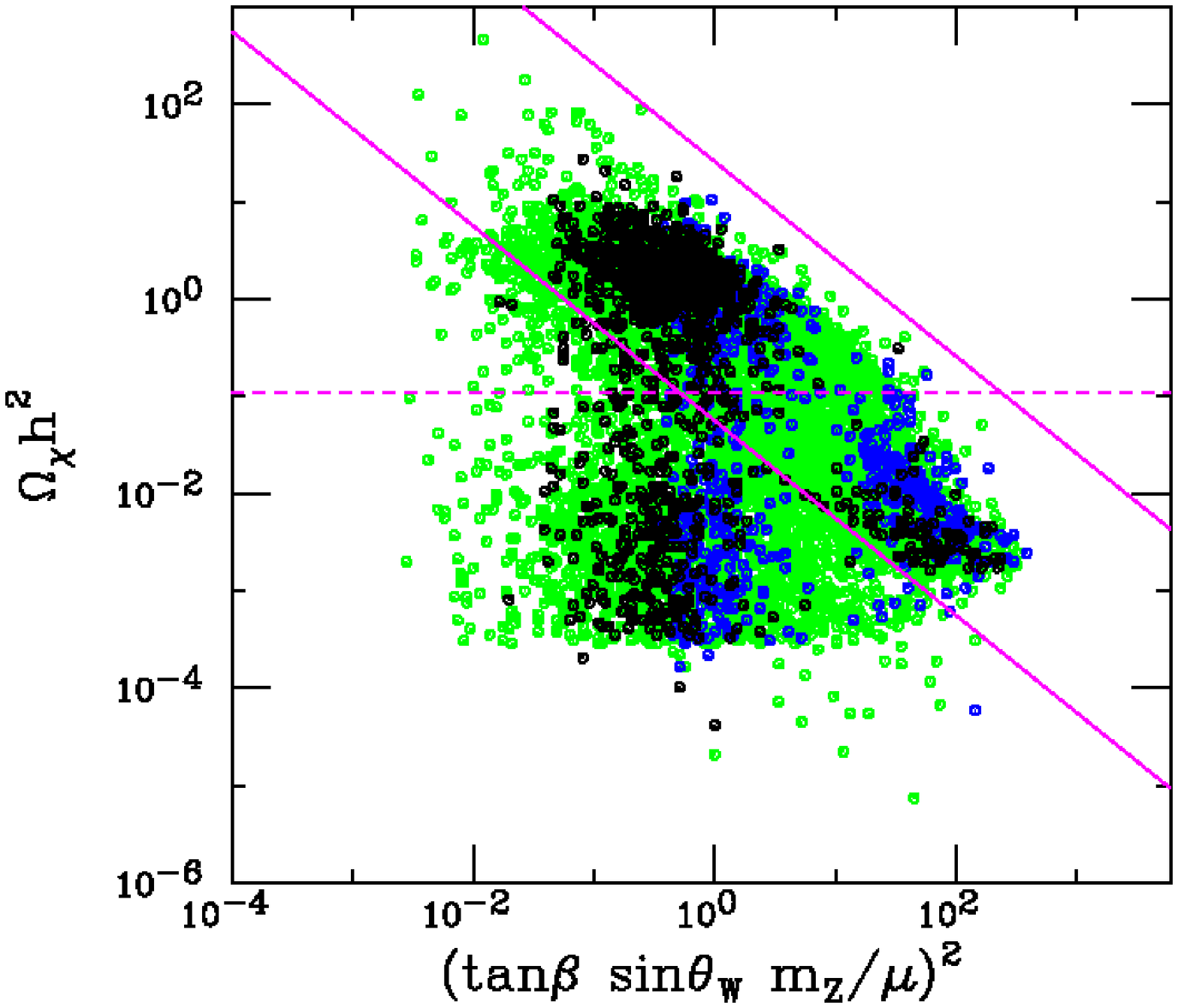}
\caption{The thermal relic abundance of neutralino dark matter. The
dashed horizontal line represents the density measured by
WMAP~\cite{wmap}. In the right frame, we compare the relic density to
the quantity $\tan^2 \beta\sin^2 \theta_W m^2_Z/\mu^2$, which for
models with a bino-like lightest neutralino is a good approximation
for $\epsilon^2_u \tan^2 \beta$, and therefore scales with the
annihilation cross section to $b \bar{b}$ via pseudoscalar Higgs
exchange.  The two solid lines represent the analytic result of this
calculation using $m_A=120$~GeV and $m_{\chi^0}=$50 (bottom) and 500
(top)~GeV.  All models shown have $95 \, \rm{GeV} < m_h < 101 \,
\rm{GeV}$, $111 \, \rm{GeV} < m_H < 119 \, \rm{GeV}$ and $0.056 \lsim
\sin^2(\beta-\alpha) \lsim 0.144$. The black points are consistent
with measurements of $B \to X_s \gamma$ at the 3$\sigma$ level and do
not violate the Tevatron constraint on the $B_s \to \mu^+ \mu^-$ branching
fraction. Blue points violate $B_s \to \mu^+ \mu^-$, but are
consistent with $B \to X_s \gamma$. Green points violate $B \to X_s
\gamma$.}
\label{fig:relic}
\end{figure}

One very attractive feature of R-parity conserving supersymmetry is
that it naturally provides a stable particle which, in many models,
can be a viable candidate for dark matter. The lightest neutralino is
particularly appealing in this respect~\cite{dmreview}.

Neutralinos can annihilate through a variety of channels, including
through the exchange of CP-even or odd Higgs bosons, charginos,
neutralinos, sfermions and gauge bosons. Which annihilation channel(s)
dominates varies from model to model. In the scenario we are studying
here, however, the presence of light Higgs bosons suggest that
s-channel Higgs exchange to fermion pairs is likely to be a
particularly efficient annihilation channel, especially in those
models with moderate to large values of $\tan \beta$. The formulae for
these annihilation channels are collected in the appendix.

The annihilation cross section can be used to calculate the thermal
relic abundance present today:
\begin{equation}
\Omega_{\chi^0} h^2 \approx \frac{10^9}{M_{\rm{Pl}}} \;
                            \frac{x_{\rm{FO}}}{\sqrt{g_{\star}}} \;
                            \frac{1}{(a + 3 b/x_{\rm{FO}})},
\end{equation}
where $g_{\star}$ is the number of relativistic degrees of freedom
available at freeze out, $a$ and $b$ are the amplitudes given in the
appendix, and $x_{\rm{FO}}$ is the value at freeze out:
\begin{equation}
x_{\rm{FO}} \approx \ln \left( \sqrt{\frac{45}{8}} \; 
                               \frac{m_{\chi^0} M_{\rm{Pl}} (a+6b/x_{\rm{FO}})}
                                    {\pi^3  \sqrt{g_{\star} x_{\rm{FO}}}}
                        \right).
\end{equation}

Over the range of cross sections and masses expected for the lightest neutralino, $x_{\rm{FO}} \approx 20-30$. To yield the density of cold dark matter
measured by WMAP~\cite{wmap}, we thus require $a + 3b/x_{\rm{FO}} \approx
3 \times 10^{-26}$ cm$^3$/s. The main contribution to $a$ and $b$ are s-channel CP-odd and CP-even Higgs exchange, respectively. From the amplitudes given in the appendix, we find $a \sim b \times \tan^2 \beta/3$, so we conclude that the $b$-term induced through CP-even Higgs plays a subdominant role in the relic density calculation to the CP-odd Higgs induced $a$-term. In the left frame of
Fig.~\ref{fig:relic}, we plot the relic density of the lightest
neutralino versus its mass. We have calculated this quantity using the
micrOMEGAs program~\cite{micromegas}.\smallskip

Over the vast majority of supersymmetric parameter space, the
composition of the lightest neutralino is dominated by its bino
component, although a small admixture of wino or higgsino is
common. For a bino-like neutralino with large or moderate $\tan
\beta$, the annihilation cross section through pseudoscalar Higgs
exchange scales as $\epsilon^2_u \tan^2 \beta$. This up-type higgsino
component can be approximated in most models by $\epsilon^2_u \approx
\epsilon^2_B \sin^2 \theta_W \sin^2 \beta \, m^2_Z/\mu^2$. With this
in mind we plot the neutralino relic density versus the quantity
$\tan^2 \beta\sin^2 \theta_W m^2_Z/\mu^2$ in the right frame of
Fig.~\ref{fig:relic}. The two solid lines shown are the analytic
results for $m_A=120$~GeV and $m_{\chi^0}=$50 (bottom) and 500 (top)
GeV in this approximation. The majority of models we find fall within
this range, suggesting that their annihilation is in fact dominated by
and s-channel pseudoscalar Higgs exchange to $b \bar{b}$. As
expected, no models fall above this range. In the models which lie
below this range, another annihilation mode or modes (such as
t-channel sfermion exchange, s-channel $Z$ exchange or s-channel
pseudoscalar Higgs exchange to top quark pairs) must contribute
substantially, thus lowering the relic abundance
accordingly. Coannihilations between the lightest neutralino and
another superpartner may also reduce the relic density.

From the right frame of Fig.~\ref{fig:relic}, we can infer that for
those models which generate the measured relic abundance (those which
fall along the horizontal dashed line), most appear to annihilate
substantially through pseudoscalar Higgs exchange to $b
\bar{b}$. Models with a bino-like neutralino with a small higgsino
admixture and moderate to high value of $\tan \beta$ are often capable
of generating the measured density of dark matter.\medskip

\begin{figure}[t]
\includegraphics[width=2.0in,angle=90]{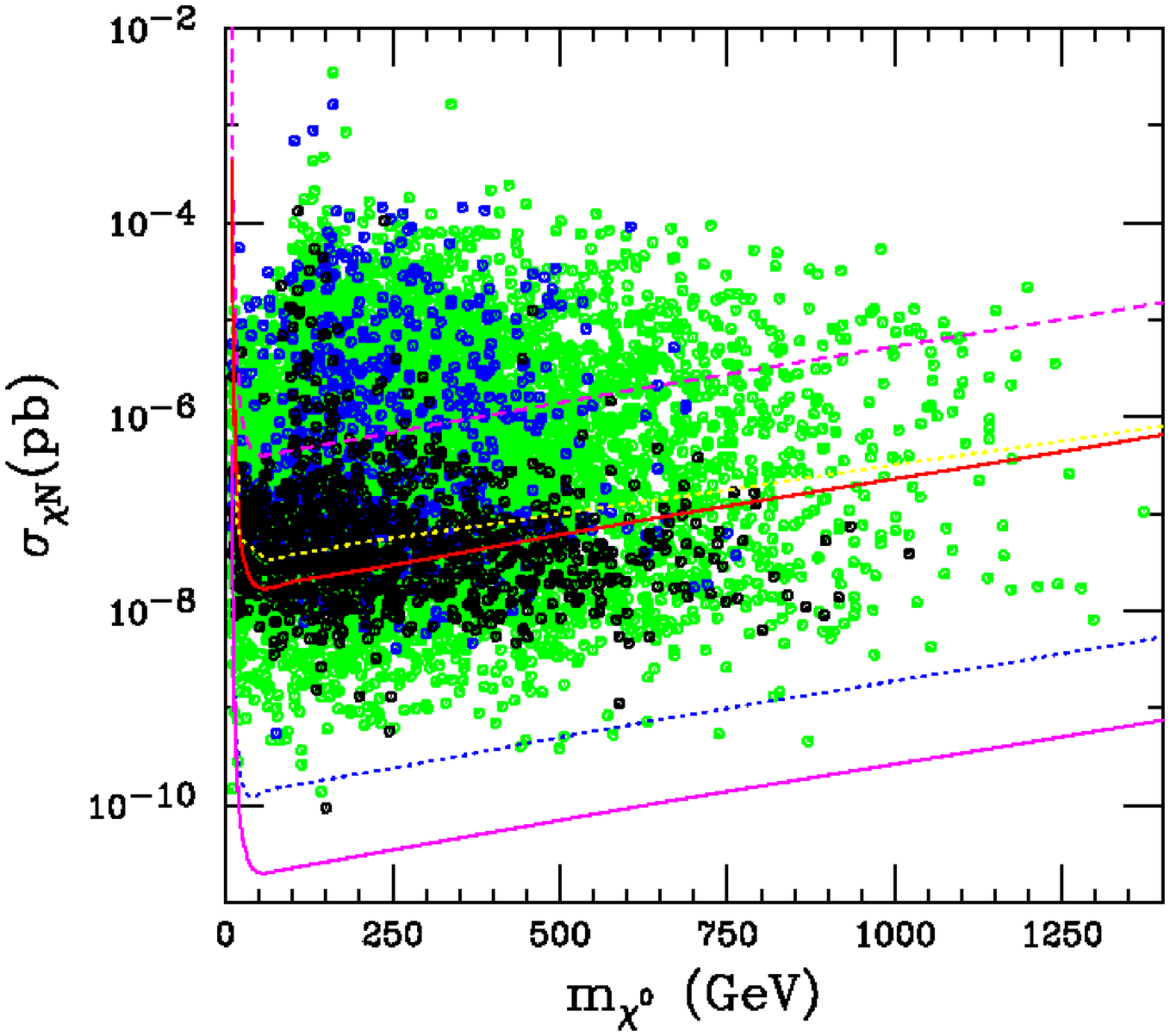} \hspace*{30mm}
\includegraphics[width=2.0in,angle=90]{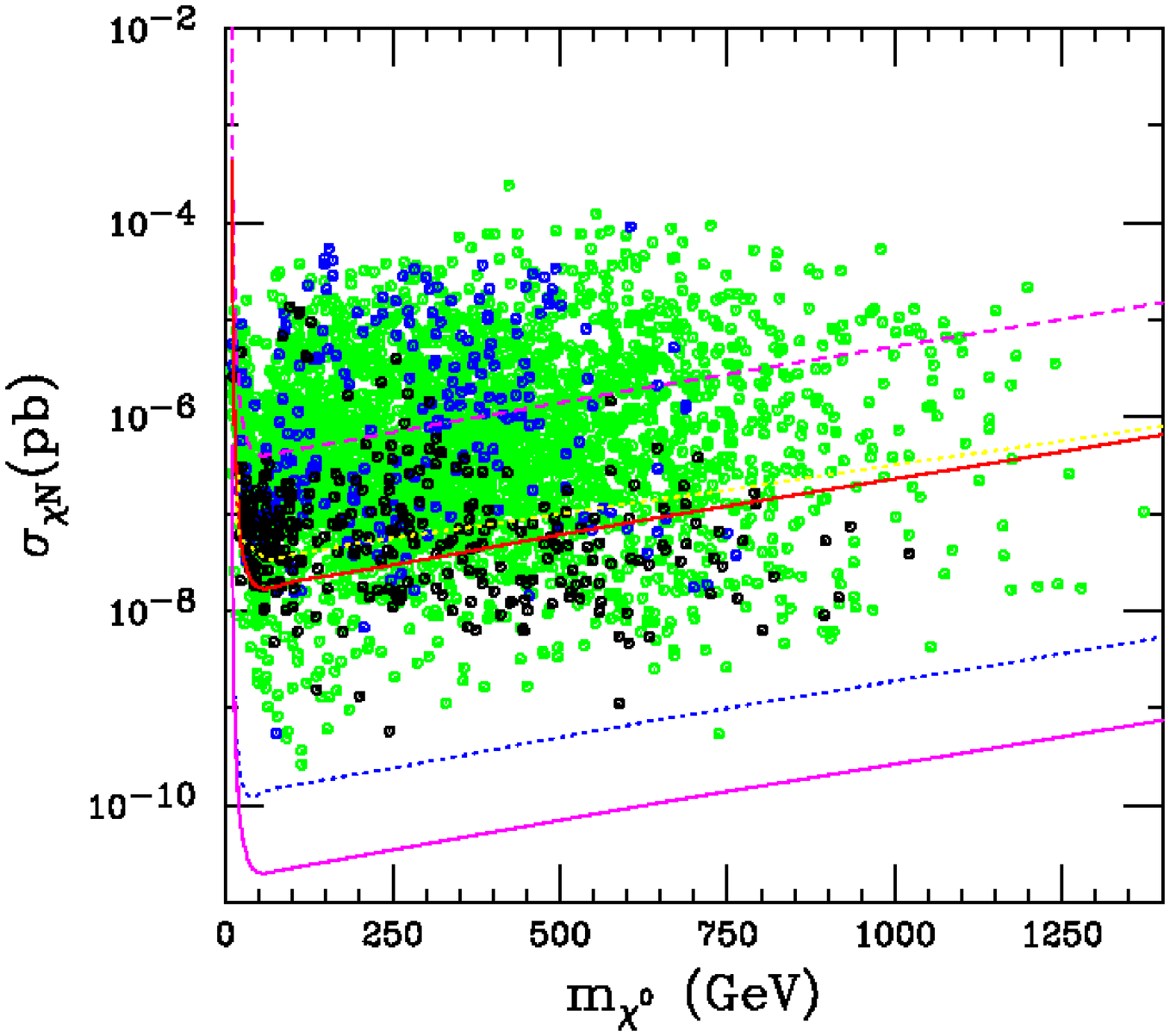}
\caption{The spin-independent neutralino-nucleon elastic scattering
cross section is shown versus the neutralino mass. All points shown
have $95 \, \rm{GeV} < m_h < 101 \, \rm{GeV}$, $111 \, \rm{GeV} < m_H
< 119 \, \rm{GeV}$ and $0.056 \lsim \sin^2(\beta-\alpha) \lsim
0.144$. The left frame includes models with any relic density, while
those models shown in the right frame generate an abundance of dark
matter within a factor of ten of the observed quantity. The dashed
curve is the current limit placed by the CDMS experiment. The dotted
yellow, solid red, dotted blue and solid magenta curves (from top to
bottom) are the approximate projected limits of GERDA, CDMS,
ZEPLIN-MAX and Super-CDMS (phase III), respectively.  The black points
are consistent with measurements of $B \to X_s \gamma$ at the
3$\sigma$ level and do not violate the Tevatron constraint on the $B_s \to
\mu^+ \mu^-$ branching fraction. Blue points violate $B_s \to \mu^+
\mu^-$, but are consistent with $B \to X_s \gamma$. Green points
violate $B \to X_s \gamma$.}
\label{fig:direct}
\end{figure}

The prospects for the direct detection of particle dark matter are
quite encouraging for a neutralino in conjunction with a light Higgs
sector. In many of the models we study here, the spin-independent
neutralino-nucleon elastic scattering cross section is dominated by
the t-channel exchange of CP-even Higgs bosons. The cross section for
this process is roughly given by
\begin{equation}
\sigma_{\chi N} \approx \frac{8 G_F^2 m_Z^2}{\pi m_H^4} 
                        \left( \frac{m_p m_{\chi^0}}
                                    {m_p+ m_{\chi^0}}
                        \right)^2  \, 
                        \left( -\epsilon_B \sin\theta_W 
                               +\epsilon_W \cos \theta_W
                        \right)^2 \, 
                        \epsilon_H^2  
                        \left[ \sum_q  \frac{m_q}{\cos \beta} 
                               \langle N|q \bar{q}|N \rangle 
                        \right]^2,
\end{equation}
where the sum is over quark types and $\langle N|q \bar{q}|N \rangle$
are the matrix elements over the nucleonic state. The quantity $\cos
\beta$ should be replaced with $\sin \beta$ for up-type quarks.
$\epsilon^2_B$, $\epsilon^2_W$ and $\epsilon^2_H$ are the bino, wino
and higgsino fractions of the neutralino. In the sum over quark
species, the strange quark contribution dominates with $m_s \langle
N|s \bar{s}|N \rangle \approx 0.2$~GeV.  For models with $\sim100$~GeV
Higgs bosons and a bino-like LSP, we estimate
\begin{equation}
\sigma_{\chi N} \sim 10^{-7}-10^{-6}\, \rm{pb} \,\, 
                     \left( \frac{\epsilon_H^2}{0.01} \right) \,  
                     \left(\frac{\tan \beta}{20} \right)^2.
\end{equation}

Of course other elastic scattering channels (squark exchange in
particular) can also contribute substantially. In
Fig.~\ref{fig:direct} we show the spin-independent
neutralino-nucleon elastic scattering cross section for models in this
scenario. Models shown in the right frame each generate an abundance
of neutralino dark matter within one order of magnitude of the
observed density, while those models shown in the left frame may or
may not. The dashed curve is the current limit placed by the CDMS II
experiment~\cite{cdms}. The dotted yellow, solid red, dotted blue and
solid magenta curves (from top to bottom) are the approximate
projected limits of the GERDA~\cite{gerda}, CDMS II,
ZEPLIN-MAX~\cite{zeplinmax} and Super-CDMS (phase III) experiments,
respectively. Edelweiss~\cite{edelweiss} should also reach a
sensitivity similar to that of CDMS. To calculate these elastic
scattering cross sections, we have used the DarkSusy
program~\cite{darksusy}.\smallskip

From Fig.~\ref{fig:direct}, it is obvious that the prospects for
direct detection are excellent for this class of supersymmetric models
with light Higgs sectors which often dominate the cross section. Most of the favored models are within the reach of the
CDMS and Edelweiss experiments and those remaining models below this level of
sensitivity should be within the reach of next generation experiments
such as ZEPLIN-MAX or Super-CDMS. 

It is straight forward to understand why models with very small elastic scattering cross sections do not appear in Fig.~\ref{fig:direct}. If we limit the magnitude of $\mu$ to be less than a few TeV for fine tuning reasons, the higgsino fraction of the lightest neutralino is then bounded from below to be $\epsilon^2_u \gsim \sin^2 \theta_W \, \sin^2 \beta \, m^2_Z/(3\,\rm{TeV})^2 \sim 0.0002$. From this, we see that $\sigma_{\chi N}$ cannot be smaller than $10^{-9}$ to $10^{-10}$~pb without $|\mu|$ being unnaturally large.

In addition to these excellent prospects for direct dark matter searches,
the characteristics of the lightest neutralino in this class of models
are fairly ideal for the purposes of indirect detection. Since the
annihilation cross section during freeze out is dominated by the first
term in the expansion, $\langle \sigma v\rangle = a + b x +
\mathcal{O}(x^2)$, the annihilation cross section relevant for
neutralinos annihilating near the galactic center, throughout the
galactic halo, or elsewhere of interest to indirect detection, is the
maximum value consistent with the measured relic abundance, $\langle
\sigma v\rangle \approx 3 \times 10^{-26}$ cm$^3$/s. This makes it
rather likely that such models with be within the reach of future indirect
detection experiments searching for neutralino annihilation products in the form of gamma-rays~\cite{gammaraysdark} or anti-matter~\cite{positrons}. The
prospects for detecting such a neutralino in future cosmic
positron experiments are especially promising.\bigskip

\section{Light MSSM Higgs Bosons at the LHC}
\label{sec:coll}

Before we begin discussing the prospects of observing a light MSSM Higgs
sector at the LHC, we should point out that it has been known for a long time
that one CP-even Higgs scalar in the MSSM is guaranteed to be seen in
weak-boson-fusion production with a subsequent decay $h,H \to
\tau\tau$~\cite{no_lose_lhc}. The search strategy is identical to the Standard
Model search for low Higgs masses and has been extensively studied including
detector effects~\cite{atlas_wbf}. In the special case of three light MSSM
Higgs scalars, this discovery channel is known to face a challenge: the
$\tau\tau$ mass resolution will not be sufficient to resolve the two CP-even
scalars if their mass difference is less than ${\cal O}(5)$~GeV --- they
will appear as one sightly wider Higgs resonance. Moreover, the pseudoscalar
Higgs only couples to gauge bosons through a dimension-5
operator~\cite{lhc_wcoup}, thus it will not be produced in weak boson fusion.
The case of three almost mass-degenerate light Higgs scalars has been
specifically studied, with an emphasis on distinguishing the three mass peaks
in the decay to muons~\cite{eduard}. The conclusion is that for Higgs masses
below 140~GeV, the inclusive Higgs search is challenging and that the heavy
Higgs bosons will probably not be separately observable in this channel. The
more promising strategy is to look for bottom-Higgs associated production with
a subsequent Higgs decay to muons and at least one $b$ tag. The Yukawa
coupling for the heavy scalar and the pseudoscalar Higgs can then be
$\tan\beta$ enhanced and the mass peaks might be resolved for $\tan\beta
\gtrsim 30$ and $m_A \gtrsim 130$~GeV.\smallskip

\begin{figure}[t]
\includegraphics[width=2.0in,angle=90]{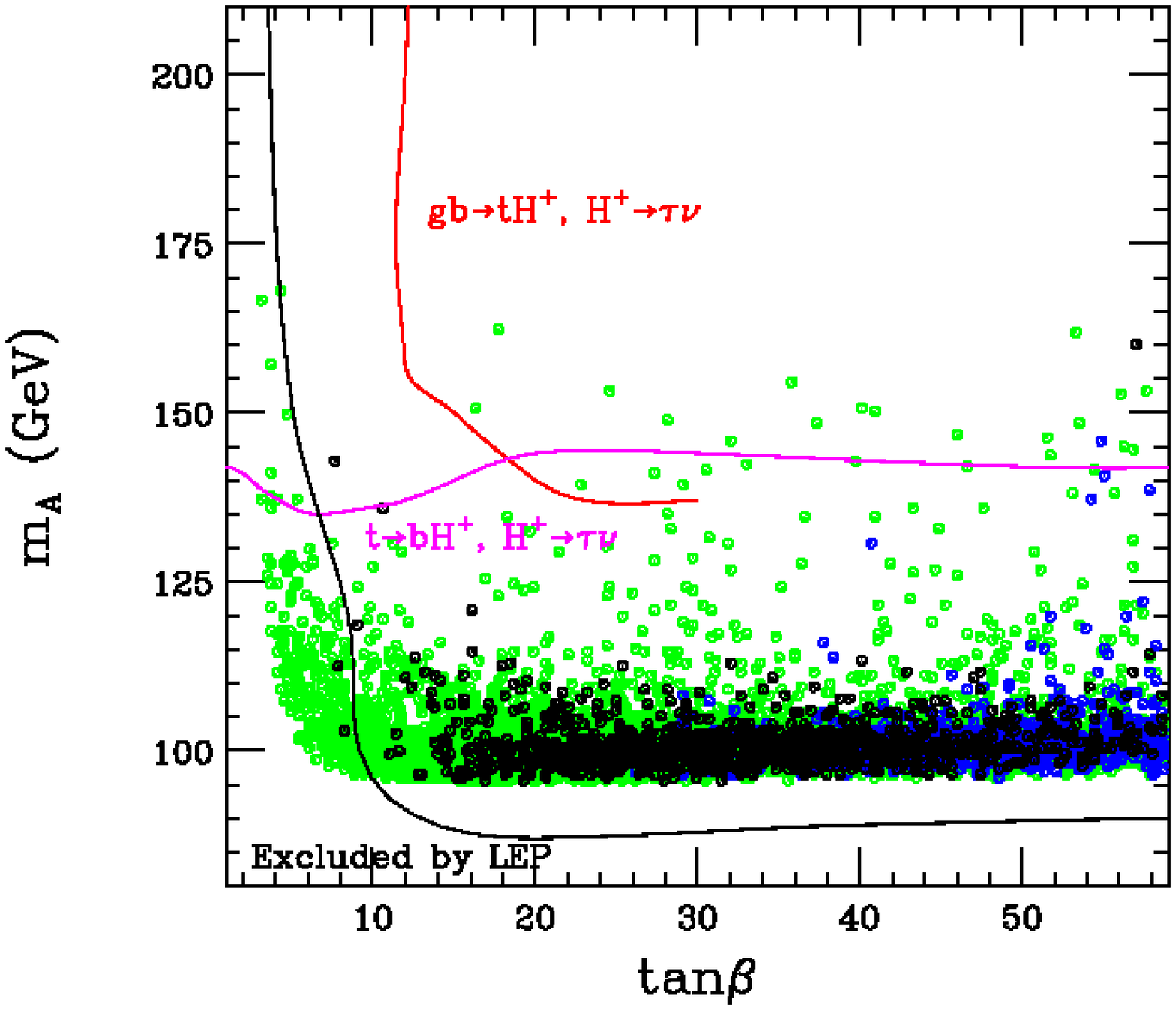} \hspace*{30mm}
\includegraphics[width=2.0in,angle=90]{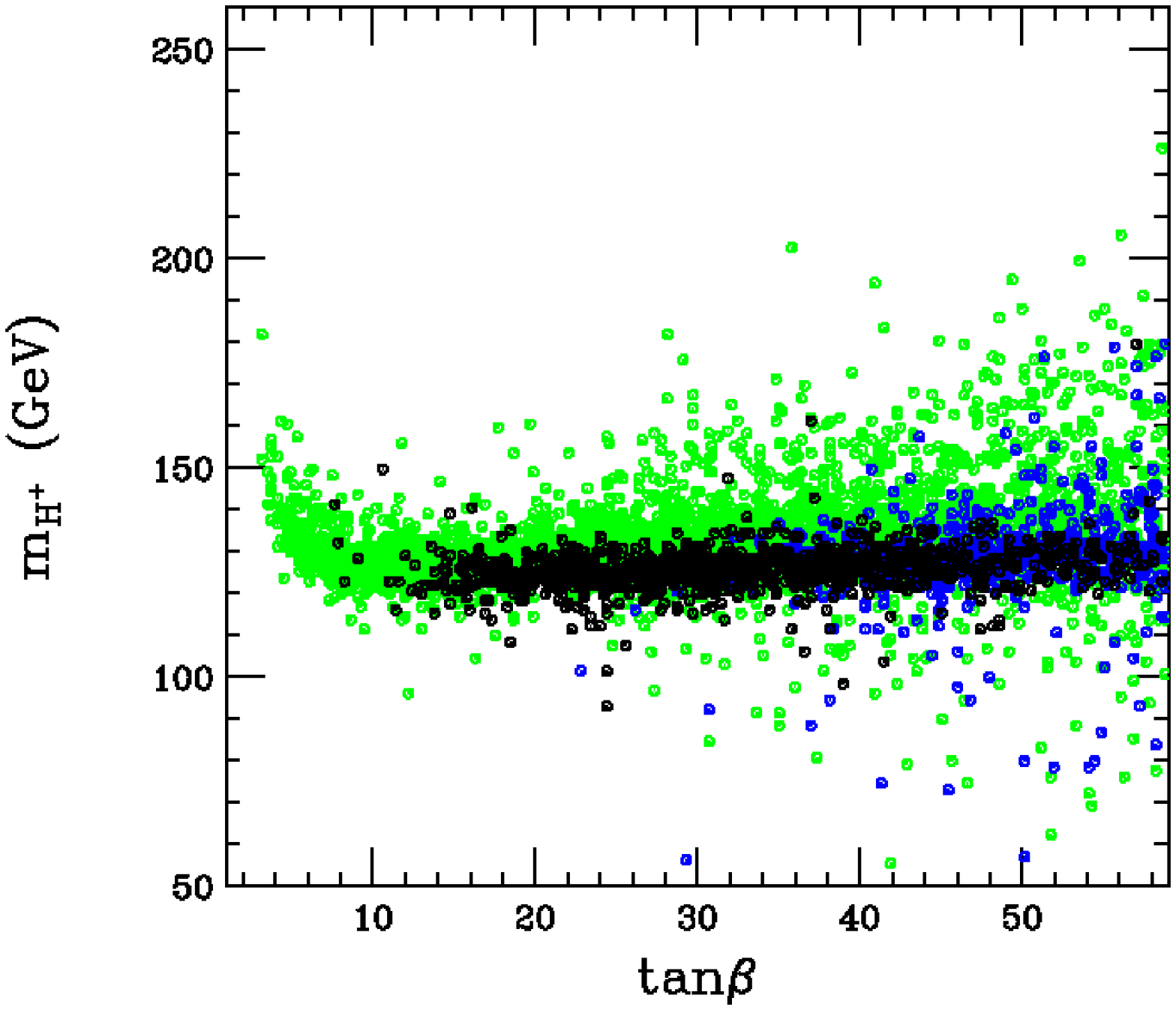}
\caption{Left: The ($5\sigma$, 300 fb$^{-1}$) discovery potential for a
charged Higgs boson at the LHC. Below the nearly horizontal (magenta)
line, charged Higgs bosons can be identified through the $t \to
H^{\pm} b$, $H^{\pm} \to \tau \nu$ channel. Models falling to the
upper right of the red line can be discovered through the channel $b g
\to t H^{\pm}$, $H^{\pm} \to \tau \nu$. Models with very light $A$,
$H^{\pm}$ or very small $\tan \beta$ are excluded by LEP. Right: the same points shown in the $m_{H^+}$ vs $\tan \beta$ plane. All models
shown have $95 \, \rm{GeV} < m_h < 101 \, \rm{GeV}$, $111 \, \rm{GeV}
< m_H < 119 \, \rm{GeV}$ and $0.056 \lsim \sin^2(\beta-\alpha) \lsim
0.144$. The black points are consistent with measurements of $B \to
X_s \gamma$ at the 3$\sigma$ level and do not violate the Tevatron
constraint on the $B_s \to \mu^+ \mu^-$ branching fraction. Blue
points violate $B_s \to \mu^+ \mu^-$, but are consistent with $B
\to X_s \gamma$. Green points violate $B \to X_s \gamma$.}
\label{fig:lhc1}
\end{figure}

The general lesson we learn from detailed studies by ATLAS and CMS is that it
might well be easier to discover a charged Higgs boson than the additional
neutral scalars. A charged Higgs can be searched for either in anomalous top
decays, or it can
be directly produced as $bg \to tH^-$. In both cases the decay to a tau
lepton is most promising~\cite{charged_lhc}. The coverage of models in the ($m_A$ -- $\tan\beta$) plane from this channel has historically included a hole for charged Higgs
masses just above the top mass. We emphasize that this hole does not mean that
the charged Higgs will be missed at the LHC in this parameter range. It just
means that we will have to combine the search strategies for (off-shell)
anomalous top decays with the usual associated
production~\cite{light_charged}. For the purpose of this paper we will use a
preliminary parton level analysis~\cite{moretti}. We include this result in Fig.~\ref{fig:lhc1}

\subsection{Light charged Higgs bosons}

Phenomenologically speaking, the easiest searches for charged Higgs bosons are in the mass
range $\mhp \lsim m_t-m_b$. A light charged Higgs will almost exclusively decay to a tau lepton, because the decays to $Wh^0$ and $t\bar{b}$ are kinematically
closed in our scenario. This means we can look for top decays to $b \ell \bar{\nu}$, where
there are too many final state taus as compared to muons and electrons. In top
pair production we let the second top decay into $Wb$, where the $W$ can decay
leptonically and hadronically (the latter channel dominates the reach in the
charged Higgs mass)~\cite{topdec_theo}. The tau from the charged Higgs
typically forms a jet. The partial width of a top decay to a charged Higgs
contains two terms: $m_b^2 \tan^2\beta + m_t^2/\tan^2\beta$. Hence, there will
be large branching fractions both for small and large values of
$\tan\beta$ with the weakest point around $\tan\beta \sim \sqrt{m_t/m_b} \sim
7$. Because at the LHC these searches are sensitive to anomalous branching
fractions of the order of $1\%$~\cite{topdec_lhc}, we can expect them to probe
charged Higgs masses up to $\mhp \lsim m_t-m_b \sim 160$~GeV. Including
off-shell effects, this reach might even be extended by a few~GeV. At the
Tevatron the same searches are currently in progress, but are
strongly statistics limited~\cite{topdec_tev}. They only cover anomalous
branching fractions of the order of $60\%$, corresponding to an enhanced
Yukawa coupling $\tan\beta \gtrsim 20$ and masses $\mhp \lsim 140$~GeV.

In Fig.~\ref{fig:lhc1} we overlay our MSSM parameter points with the
LHC (ATLAS) $5\sigma$ reach. The magenta line corresponds to the
low-luminosity results from anomalous top decays. The slight bending comes
from the mass relation between $m_A$ and $\mhp$ as well as from the
coupling suppression for intermediate $\tan\beta$. We see that most of
the parameter points with two light Higgs scalars will be clearly
visible in anomalous top decays at the LHC, possibly even at the
Tevatron.\medskip

\subsection{\boldmath Heavy charged Higgs bosons with large $\tan\beta$}

The best-studied regime for charged Higgs searches at the LHC is the region of
large $\tan\beta \gtrsim 10$ and charged Higgs masses well above the top
threshold. Following the approximation in Eq.(\ref{eq:mt_approx}), we see that
the required two light scalar Higgs bosons imply $\sqrt{m_A^2 + m_Z^2 + \epsilon}
\sim 150$~GeV. Requiring charged Higgs masses above the top threshold
translates into $m_A > 150$~GeV, or in other words $m_Z^2 \sim -\epsilon$.
This could be possible if one stop were much lighter than the top. On
the other hand, the
$\epsilon$ approximation is based on the assumption of non-mixing stops, so we
postpone this light-stop scenario to the next subsection. The other way of
enhancing the mass difference between the average CP-even Higgs mass and the
charged Higgs mass (beyond the simple $m_t^4$ approximation) and to escape the
anomalous top decays is to create 
large bottom Yukawa corrections through large values of $\tan\beta$ (while keeping
the $\mu$ parameter small). This leads to additional terms contributing to the
charged Higgs mass squared, the leading terms of which are proportional to
$m_t^2 m_b^2$. The two kinds of contributions which appear with this mixed
Yukawa coupling are either proportional to
$\log (\mst{1} \mst{2}/m_t^2)$, or they come without this logarithm and
are directly proportional to the average stop mass instead~\cite{carlos}. We
see their effect in
Fig.~\ref{fig:lhc1} when we compare the $\tan\beta$ dependence of $m_A$ and
$\mhp$. For all value of $\tan\beta$ the requirement of two light Higgs scalar
limits the pseudoscalar mass to a narrow corridor. We find hardly any SUSY
scenarios with $m_A > 110$~GeV. By considering large values of $\tan\beta$, we
can increase the charged Higgs mass by 50~GeV and avoid the top decay threshold.\smallskip

Because large values of $\tan\beta$ are required to at the same time
have two light Higgs scalars and a heavier charged Higgs, we are automatically
driven into the region where the charged Higgs can be found in
$tH^-$ production at the LHC. The only issue is the hole which usually appears
in the LHC coverage region for charged Higgs masses between 160 and 200~GeV.
As mentioned above, this hole is an artifact of the two search strategies
meeting in this mass range. The associated production process and the
anomalous (off-shell) top decay have to be combined to cover this hole. This
kind of study is ongoing in CMS and in ATLAS~\cite{moretti} and we do not
expect this to be a problem once these analyses are actually performed on data. In
Fig.~\ref{fig:lhc1} the densities of points suggests that it is very unlikely
to find such a light-Higgs MSSM scenario where the LHC sees only one neutral Higgs scalar
and no charged Higgs
(we find only one point somewhat close to this region in our scan). However,
we will study these kinds of parameter points and their discovery prospects at
the LHC in more detail next, to ensure there is no hole in the LHC discovery range.

\subsection{Heavy charged Higgs bosons and light stops}
\label{sec:weird}

\begin{figure}[t]
\includegraphics[width=2.0in,angle=90]{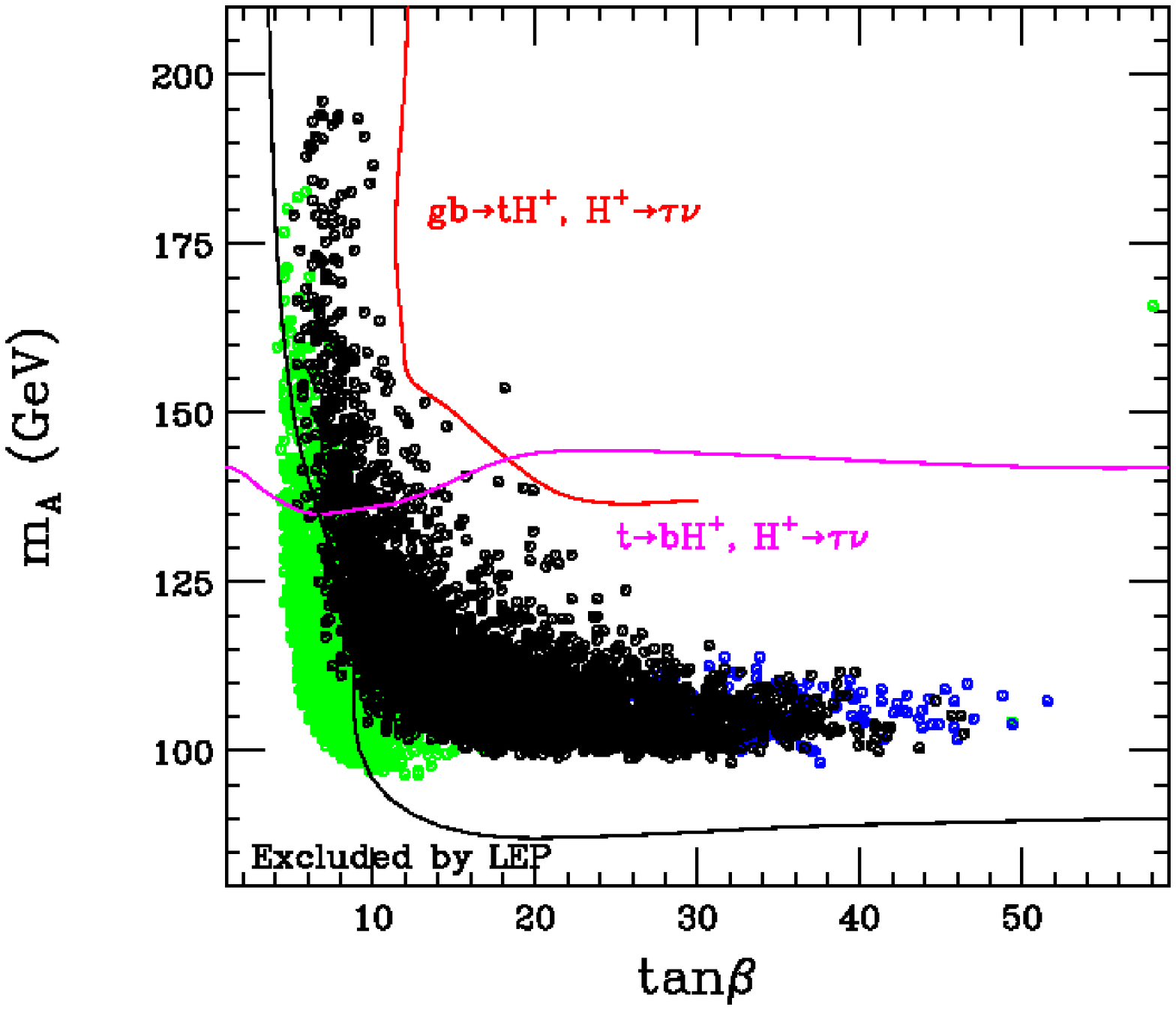} 
\includegraphics[width=2.0in,angle=90]{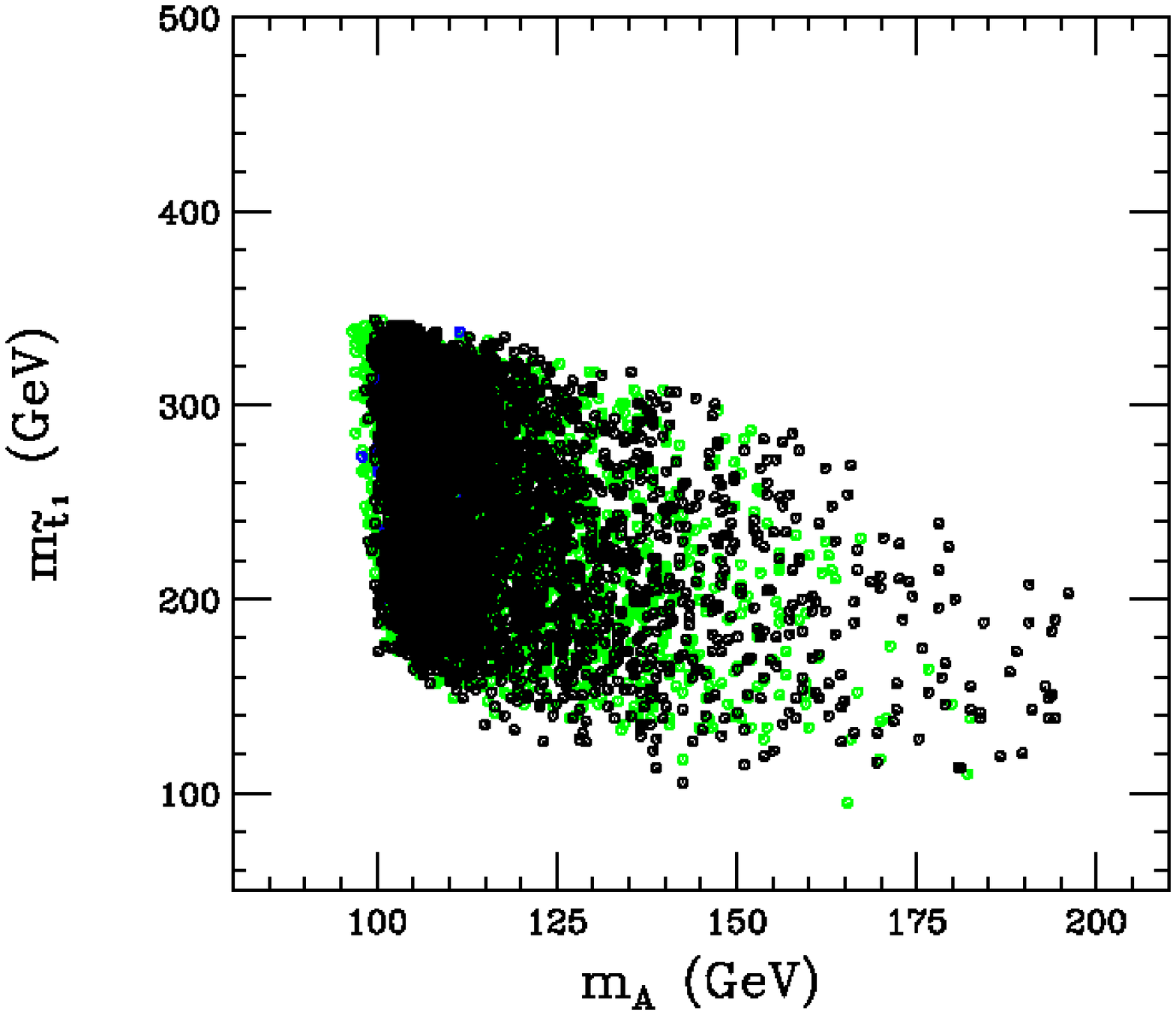}
\includegraphics[width=2.0in,angle=90]{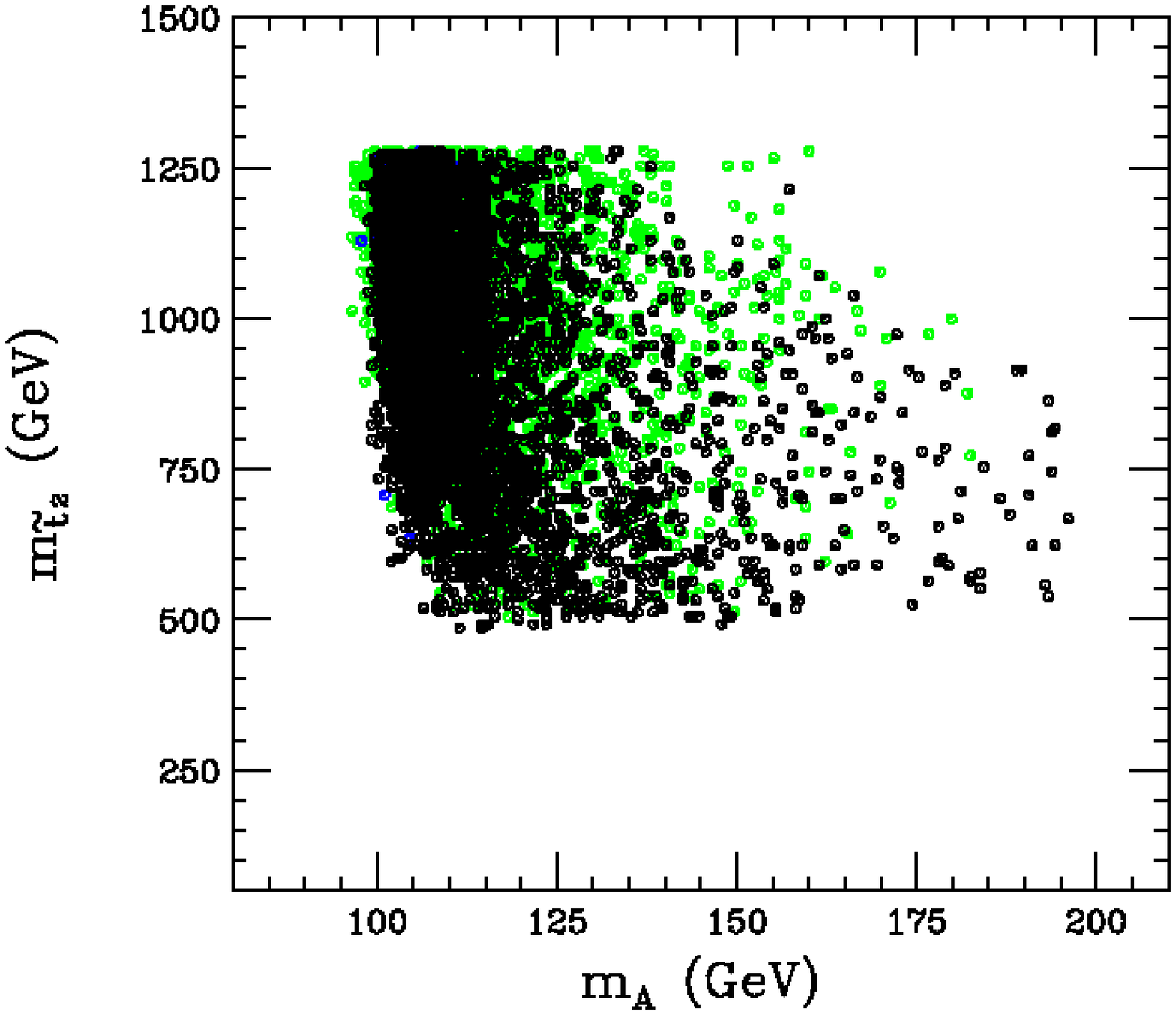}
\caption{Left: same as in Fig.~\ref{fig:lhc1}, but with parameters
varied as described in the text of section~\ref{sec:weird}. Right two panels: the
light stop masses for the same set of parameter points. All models shown
have $95 \, \rm{GeV} < m_h < 101 \, \rm{GeV}$, $111 \, \rm{GeV} < m_H
< 119 \, \rm{GeV}$ and $0.056 \lsim \sin^2(\beta-\alpha) \lsim
0.144$. The black points are consistent with measurements of $B \to
X_s \gamma$ at the 3$\sigma$ level and do not violate the Tevatron
constraint on the $B_s \to \mu^+ \mu^-$ branching fraction. Blue
points violate $B_s \to \mu^+ \mu^-$, but are consistent with $B
\to X_s \gamma$. Green points violate $B \to X_s \gamma$.}
\label{fig:lhc2}
\end{figure}

As described above, there is this remaining part of the light-scalars parameter
space where charged Higgs searches in anomalous top decays and in the
associated production with a top are challenging. It requires charged Higgs
masses just above the top threshold and small $\tan\beta$. Note
that we do not claim that LHC will not see a charged Higgs in this
region of parameter space. Instead, we emphasize that covering this hole is a
crucial task for the near future. We perform a dedicated scan
around the one point which appears in Fig.~\ref{fig:lhc1} close to the hole in
the LHC reach for intermediate $\tan\beta <20$ and $\mhp \sim 180$~GeV.
We vary each mass parameter up to 50\% above or below
the value of the single point at $\tan \beta \approx 8$, $m_A \approx 140$
shown in Fig.~\ref{fig:lhc1}. Only $\tan \beta$ we vary over the entire
range 1-60. At this specific point, $M_2$ and $M_3$ are both $\sim1$ TeV, $\mu\sim2$ TeV,  $A_t \sim -500$ GeV, and the bino mass is somewhat light, $M_1 \approx 140$ GeV.\medskip

To see what kind of MSSM parameter choice can enhance the charged Higgs mass
compared to the CP-even scalar masses, we extend the analytical approximation
of Eq.(\ref{eq:mt_approx}) to mixing stops~\cite{carlos}. To avoid the LHC
limits, we are now interested in parameter points with as small as possible
$\tan\beta$, so we can neglect the bottom Yukawa coupling. Instead, we 
take into account stop mixing. In two limits the expressions are particularly simple:
\begin{alignat}{6}
\mhp^2 -\left(m_h^2+m_H^2\right) &=&& \; m_W^2 - m_Z^2 \;
                                       + \frac{G_F m_t^4}{4 \sqrt{2} \pi^2} \;
                                         \frac{A_t^2 (A_t^2+\mu^2)}
                                              {\mst{1}^2 \mst{2}^2}
                                 \qquad \qquad \qquad &&(\tan\beta \to 1)
                                 \notag \\
\mhp^2 -\left(m_h^2+m_H^2\right) &=&& \; m_W^2 - m_Z^2 \;
                                       + \frac{G_F m_t^4}{2 \sqrt{2} \pi^2} \;
                                         \frac{(A_t-\mu)^2 (A_t^2+\mu^2)}
                                              {\mst{1}^2 \mst{2}^2}
                                               &&(\tan\beta \gg 1).
\label{eq:mt_mix}
\end{alignat}
These results point at another way of increasing the charged Higgs
mass while keeping two light Higgs scalars: we choose fairly light stops
and increase the trilinear coupling, $A_t$, or the $\mu$ parameter. We have
checked explicitly that for large values of $\tan\beta$ the same can be
achieved through increasing $A_b$. However, keeping both stops light and at the
same time increasing one way or another the stop mixing parameters, $A_t -
\mu/\tan\beta$, means that at least the lighter stop and the lighter sbottom will be very light.
  This is precisely what we observe in Fig.~\ref{fig:lhc2}, where we see that the lightest stop mass is always lighter than approximately 340~GeV.\smallskip

Depending on the neutralino and chargino masses there are two possible stop
decays we have to consider when we search for light stops at the Tevatron and
at the LHC. If the stop is heavier than the lightest chargino, it will decay mostly into $b \tilde{\chi}_1^+$. If this decay channel is kinematically
forbidden, as the LEP2 limits suggest for the Run I stop searches at the
Tevatron, the stop has to decay to $q \tilde{\chi}_1^0$.

The production cross sections for light stops are large: at the Tevatron, a
200~GeV stop will be produced with a total cross section of approximately 300~fb, and at the
LHC a 300~GeV stop will have a production rate around 10~pb. This means that at the LHC, although this kind of MSSM parameter point might be hard to find a
charged Higgs boson, there will be huge numbers of top squarks
establishing supersymmetry. Ongoing Tevatron searches with a projected Run~II
reach of well above 200~GeV (dependent on the neutralino and chargino masses)
will already severely constrain models with light stops.\medskip

\begin{figure}[t]
\includegraphics[width=2.0in,angle=90]{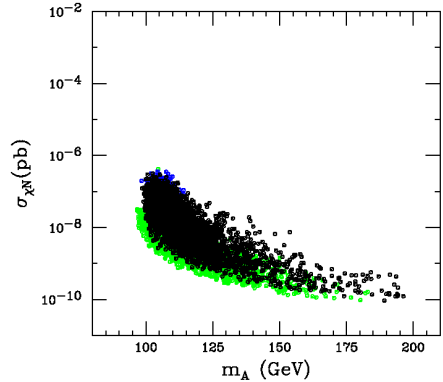} 
\caption{The spin-independent neutralino-nucleon elastic scattering
cross section is shown versus the CP-odd Higgs mass for the range of
models described in section~\ref{sec:weird} of the text. For those models
in which charged Higgs searches at the LHC will be difficult at the
LHC ($m_A \gsim 140$~GeV), this cross section is rather small,
$\sigma_{\chi N} \sim 10^{-9}-10^{-10}$~pb. These models each contain
a rather light neutralino, however, $m_{\chi^0} \approx 70-200$~GeV,
making most of them accessible to ZEPLIN-MAX and all of them
accessible to Super-CDMS (see Fig.~\ref{fig:direct}). All models
shown have $95 \, \rm{GeV} < m_h < 101 \, \rm{GeV}$, $111 \, \rm{GeV}
< m_H < 119 \, \rm{GeV}$ and $0.056 \lsim \sin^2(\beta-\alpha) \lsim
0.144$. The black points are consistent with measurements of $B \to
X_s \gamma$ at the 3$\sigma$ level and do not violate the Tevatron
constraint on the $B_s \to \mu^+ \mu^-$ branching fraction. Blue
points violate $B_s \to \mu^+ \mu^-$, but are consistent with $B
\to X_s \gamma$. Green points violate $B \to X_s \gamma$.}
\label{directweird}
\end{figure}

One might worry that in these models with lower $\tan \beta$ direct dark matter detection might be more difficult. In Fig.~\ref{directweird} we show that this is in fact
the case, with spin-independent elastic scattering cross sections as
low as $10^{-10}$~pb. In each of these models, however, there is a
rather light neutralino, $m_{\chi^0} \lsim 200$~GeV, which makes
all of these models observable by Super-CDMS and many of them
observable by ZEPLIN-MAX, even for $\sigma_{\chi N} \approx 10^{-10}$
pb (see Fig.~\ref{fig:direct}).

\section{Conclusions}

In this article, we describe the phenomenology of models within the MSSM
containing CP-even Higgs bosons with masses of approximately 98 and 114~GeV,
motivated by the excesses observed by LEP~\cite{drees}. We study in detail
indirect constraints on such light-Higgs models, coming from $B \to
X_s\gamma$, $(g-2)_\mu$ and $B_s \to \mu\mu$. The first two constraints
require fairly large $\tan\beta$ when compared with data, the latter tends to
disfavor this regime, and the upcoming Tevatron Run~II results are going to
close in on our light-Higgs parameter space. However, it is possible to avoid
all indirect constraints by tuning the different weak-scale SUSY breaking
parameters.\smallskip

We find that the kind models we are interested in typically include a light
charged Higgs boson along with the other light Higgs bosons. Such a state is
observable in anomalous top decays at the Tevatron or at the LHC. Scenarios
with charged Higgs above the top threshold are automatically driven in the
larger $\tan\beta$ regime, where the charged Higgs can be detected in
bottom--gluon fusion at the LHC.  We find exceptionally models which fall
between these two mass regions, so that neither of the two searches is
optimized and a combination of the two search tactics would be needed.
However, these challenging scenarios require one very light top squark
($m_{\tilde{t}_1} \lsim 300$~GeV), so a large fraction of them will already be
ruled out at the Tevatron.\smallskip

We also explore the phenomenology for neutralino dark matter in this
class of models. The prospects for direct detection are excellent, with the
majority of models within this scenario being testable in currently operating
experiments, such as CDMS-II. We find that {\it all} models within this class
will be testable in next generation direct dark matter detection experiments,
such as ZEPLIN-MAX or Super-CDMS. The prospects for indirect are also
favorable in this scenario.

\vspace{0.5cm}

{\it Acknowledgements}: We would like to thank Prof.~Gudrun Hiller for
discussing in detail the indirect constraints from $B$ physics and for
being of great help. We would like to thank Thomas Hahn and Sven
Heinemeyer for their help with FeynHiggs. DH is supported by the
Leverhulme Trust. \vskip -0.5cm

\section{Appendix: Neutralino Annihilation Cross Sections}
The squared amplitudes for the processes, $\chi^0 \chi^0 \to A \to f
\bar{f}$ and $\chi^0 \chi^0 \to H \to f \bar{f}$, averaged over the
final state angle are given by~\cite{amp}
\begin{eqnarray}
\omega^A_{f\bar{f}} &=& \frac{C^2_{ffA} \, C^2_{\chi^0 \chi^0 A}}
                             {(s-m^2_A)^2 +  m^2_A \Gamma^2_A} \, 
                        \frac{s^2}{16 \pi} \sqrt{1 + \frac{4 m^2_f}{s}}, \\
\omega^H_{f\bar{f}} &=& \frac{C^2_{ffH} \, C^2_{\chi^0 \chi^0 H}}
                             {(s-m^2_H)^2 +  m^2_H \Gamma^2_H} \, 
                        \frac{(s-4 m^2_{\chi^0})(s-4m^2_f)}{16 \pi} 
                        \sqrt{1 + \frac{4 m^2_f}{s}}.
\end{eqnarray}
Here the label $H$ denotes either CP-even Higgs boson. $C_{ffA}$,
$C_{\chi^0 \chi^0 A}$, $C_{ffH}$ and $C_{\chi^0 \chi^0 H}$ are the
fermion-fermion-Higgs and neutralino-neutralino-Higgs
couplings. $\Gamma_{A,H}$ are the widths of the respective Higgs
bosons.

These squared amplitudes can be used to attain the thermally averaged
annihilation cross section~\cite{falkellis}
\begin{eqnarray}
\langle \sigma v
        \rangle &=& \frac{\omega(s_0)}{m^2_{\chi^0}} 
                  - \frac{3}{m_{\chi^0}} 
                    \left[ \frac{\omega(s_0)}{m^2_{\chi^0}} 
                          - 2 \omega^{\prime}(s_0) 
                    \right]T +\mathcal{O}(T^2) \nonumber \\
                &=& \frac{1}{m^2_{\chi^0}} 
                    \left[1- \frac{3T}{m_{\chi^0}} 
                    \right] 
                    \omega(s)\bigg|_{s\to 4 m^2_{\chi^0}+6 m_{\chi^0} T}
                    +\mathcal{O}(T^2),
\end{eqnarray}
where $T$ is the temperature. Keeping terms to zeroth and first order
in $T$ is sufficient for the relic abundance calculation. Writing this
as an expansion in $x=T/m_{\chi^0}$, $\langle \sigma v\rangle = a + b
x + \mathcal{O}(x^2)$, we arrive at
\begin{eqnarray}
a_{\chi\chi \to A\to f\bar{f}} &=& \frac{g^4_2 c_f m^2_f  \tan^2 \beta}
                                        {8 \pi m^2_W} 
                                   \frac{m^2_{\chi^0} \sqrt{1-m^2_f/m^2_{\chi^0}}}
                                        {(4 m^2_{\chi^0}-m^2_A)^2 + m^2_A \Gamma^2_A} \nonumber \\ 
                &\times& \left[\epsilon_U 
                               \left( \epsilon_W-\epsilon_B \tan \theta_W
                               \right) \sin \beta 
                             - \epsilon_D 
                               \left( \epsilon_W-\epsilon_B \tan \theta_W
                               \right) \cos \beta 
                         \right]^2 \nonumber \\
b_{\chi\chi \to A \to f\bar{f}} &=& 0 \nonumber \\
a_{\chi\chi \to H \to f\bar{f}} &=& 0 \nonumber \\ 
b_{\chi\chi \to H \to f\bar{f}} &=& \frac{3 g^4_2 c_f m^2_f \sin^2 \alpha}
                                         {16 \pi m^2_W \sin^2 \beta} 
                                    \frac{(m^2_{\chi^0}-m^2_f) \sqrt{1-m^2_f/m^2_{\chi^0}}}
                                         {(4 m^2_{\chi^0}-m^2_H + i m_H \Gamma_H )^2} \nonumber \\
                 &\times& \left[\epsilon_U 
                                \left( \epsilon_W-\epsilon_B \tan \theta_W
                                \right) \cos \alpha 
                              - \epsilon_D
                                \left( \epsilon_W- \epsilon_B\tan \theta_W
                                \right) \sin \alpha 
                          \right]^2 \nonumber \\
a_{\chi\chi \to h \to f\bar{f}} &=& 0 \nonumber \\ 
b_{\chi\chi \to h \to f\bar{f}} &=& \frac{3 g^4_2 c_f m^2_f \cos^2 \alpha}
                                         {16 \pi m^2_W \sin^2 \beta} 
                                    \frac{(m^2_{\chi^0}-m^2_f) \sqrt{1-m^2_f/m^2_{\chi^0}}}
                                         {(4 m^2_{\chi^0}-m^2_h + i m_h \Gamma_h )^2} \nonumber \\
                 &\times& \left[\epsilon_U 
                                \left(\epsilon_W-\epsilon_B \tan \theta_W
                                \right) \sin \alpha 
                              - \epsilon_D
                                \left(\epsilon_W-\epsilon_B \tan \theta_W
                                \right) \cos \alpha 
                          \right]^2,
\label{eq:app}
\end{eqnarray}
where $c_f$ is a color factor, equal to 3 for quarks and 1
for leptons. The quantities $\epsilon_{B,W,U,D}$ are the components of
the lightest neutralino which are bino, wino, up-type higgsino and
down-type higgsino, respectively. These expressions assume that the
final state fermions are down-type. For up-type fermions the factor of
$\tan^2 \beta$ should be replaced by $\cot^2 \beta$.

\end{document}